\documentclass[usenatbib]{mn2e}

\usepackage{amssymb}
\usepackage{graphicx}
\usepackage{rotating}
\newcommand{\Msun}{\ensuremath{{\rm M}_{\sun}}}
\newcommand{\Lsun}{\ensuremath{{\rm L_{\sun}}}}
\newcommand{\Rsun}{\ensuremath{{\rm R_{\sun}}}}
\newcommand{\iso}[2]{\hbox{${}^{#1}{\rm #2}$}}

\title[Updated AGB yields]
{Updated stellar yields from Asymptotic Giant Branch models}
\author[Amanda I. Karakas]{A. I. Karakas$^{1}$\thanks{E-mail: akarakas@mso.anu.edu.au}\\
$^{1}$Research School of Astronomy \& Astrophysics, Mount Stromlo Observatory,
Weston Creek ACT 2611, Australia}
\begin{document}



\maketitle

\label{firstpage}

\begin{abstract}

An updated grid of stellar yields for low to intermediate-mass 
thermally-pulsing Asymptotic Giant Branch (AGB) stars are presented. 
The models cover a range in metallicity $Z = 0.02, 0.008, 0.004$,
and $0.0001$, and masses between 1$M_{\odot}$ to 6$M_{\odot}$. 
New intermediate-mass ($M\ge 3M_{\odot}$) 
$Z = 0.0001$ AGB models are also presented, 
along with a finer mass grid than used in previous studies.
The yields are computed using an updated reaction rate network 
that includes the latest NeNa and MgAl proton capture rates, 
with the main result that between $\sim 6$ to 30 times less 
Na is produced by intermediate-mass models with hot bottom
burning.
In low-mass AGB models we investigate 
the effect on the production of light elements of including
some partial mixing of protons into the intershell region 
during the deepest extent of each third dredge-up episode.
The protons are captured by the abundant $^{12}$C to form 
a $^{13}$C pocket.
The $^{13}$C pocket increases the yields of $^{19}$F, $^{23}$Na, 
the neutron-rich Mg and Si isotopes, $^{60}$Fe, and $^{31}$P. 
The increase in $^{31}$P is by factors of $\sim 4$ to 20, 
depending on the metallicity. 
Any structural changes caused by the addition of the $^{13}$C 
pocket into the He-intershell are ignored.
However, the models considered are of low mass and any such 
feedback is likely to be small.  Further study is required 
to test the accuracy of the yields from the 
partial-mixing models.
For each mass and metallicity, the yields are presented in a
tabular form suitable for use in galactic chemical evolution 
studies or for comparison to the composition of planetary 
nebulae. 

\end{abstract}


\section{Introduction}

The Asymptotic Giant Branch (AGB) phase is the last nuclear burning 
phase for stars with initial masses between $\approx 0.8\Msun$ to
8$\Msun$, where the exact limits depend on the initial 
metallicity, $Z$.  During the AGB there is
a complex interplay of nucleosynthesis and mixing that alters
the surface composition of the star. The enriched AGB envelope is 
eventually expelled into the interstellar medium (ISM) 
by a slow stellar wind thus enriching the local ISM with
the products of hydrogen and helium burning, and heavy
elements produced by the slow neutron capture process
(the $s$ process).
Hence these stars are important contributors to the 
chemical evolution of galaxies and stellar systems.
For recent reviews of AGB evolution and nucleosynthesis see
\cite{busso99} and \cite{herwig05}. 

Briefly, during the TP-AGB phase the He-burning shell becomes 
thermally unstable every $10^{4}$ years or so,
depending on the core mass. The energy from the thermal pulse 
(TP) or flash drives a convective pocket in the He-rich intershell, 
that mixes the products of He-burning nucleosynthesis within this region. 
Following a TP, the convective envelope may move inward (in mass) 
to regions previously mixed by the flash-driven convective pocket. 
This inward movement of the convective envelope is known as 
the third dredge-up (TDU), and is responsible for enriching
the surface in \iso{12}C and other products of He-burning,
as well as heavy elements produced by the $s$ process in the
He-rich intershell.  
Following the TDU, the star contracts and the H-shell is 
re-ignited, providing most of the surface luminosity for 
the next interpulse period.
In AGB stars with initial masses $\gtrsim 4\Msun$, the 
base of the convective envelope becomes hot enough to 
sustain proton-capture nucleosynthesis (hot bottom burning, 
HBB).  HBB can change the surface 
composition because the entire envelope is exposed to the hot 
burning region a few thousand times per interpulse period. 
AGB stars with HBB have short lifetimes
($\tau \lesssim 100$~Myr) and are one of the stellar sites 
proposed as the polluters of globular clusters 
\citep{cottrell81,gratton04,renzini08}, even if quantitative
problems with the models exist \citep[e.g.,][]{fenner04}.

In \citet{karakas07b} we presented AGB model data and stellar 
yields for masses between 1 and 6$\Msun$, and for metallicities 
$Z=0.02,0.008, 0.004$, and $Z = 0.0001$. The main
drawbacks of that study were that the reaction network 
dated back to 2003 and there have
been significant changes to some of the important proton
and $\alpha$-capture rates since that time. Second, the 
initial compositions for the $Z=0.008$ and 0.004 models were 
not scaled solar, but had sub-solar C, N, and O 
compositions that were thought appropriate for the Large and 
Small Magellanic Clouds. We also did not investigate the
effect of including a \iso{13}C pocket in the top of the
intershell region, which is required to produce neutrons
by the reaction \iso{13}C($\alpha,n$)\iso{16}O during 
the interpulse period \citep{straniero95}. The neutrons
released are required for the $s$ process but can also
effect the abundance of elements lighter than iron in
the He-intershell, and hence the stellar yields.
Here we include a partially mixed zone in low-mass AGB models 
of 2$\Msun$, $Z = 0.0001$, and 3$\Msun$, $Z=0.02$, 0.008, and 
0.004. 

The main aims of this paper are to provide an update to the
stellar yields presented in \citet{karakas07b} using newer
reaction rates, and to provide yields calculated from
scaled-solar initial abundances for the
$Z=0.008$ and 0.004 models. We also present yields covering 
a finer mass grid than used previously, and new 
intermediate-mass $Z=0.0001$ AGB models. Here we define
intermediate-mass models as having $M > 3\Msun$, except
at $Z = 0.0001$ where $M \ge 3\Msun$.
Furthermore, we examine the effect of a partial mixing zone 
on the stellar yields of low-mass AGB stars. The new 
results are compared to the yields presented in 
\citet{karakas07b}, and to other recent AGB yields in 
the literature 
\citep[e.g.,][]{stancliffe07,stancliffe08,cristallo09,ventura09}. 
Only comparisons between detailed AGB models are made;
comparisons to synthetic AGB models \citep[e.g.,][]{marigo01} 
was discussed in detail in \citet{izzard04b}.
We also limit our discussion to AGB models with 
[Fe/H] $\gtrsim -2.3$; see \citet{campbell08} for yields
from very low-metallicity AGB models.

\section{Numerical method} \label{numerics}

We calculate the structure first and perform detailed 
nucleosynthesis calculations afterward, using a post-processing
algorithm. The details of this procedure and the codes used to 
compute the models have been previously described in 
some detail, see for example \citet{karakas02}, 
\citet{lugaro04} and \citet{karakas07b}.  In regards
to the input physics used in the stellar structure 
computations, we employ the mixing-length theory 
of convection with $\alpha = 1.75$ \citep{frost96}. At 
high temperatures we use the OPAL opacities \citep{iglesias96},
and at low temperatures we employ two different
prescriptions. The models of \citet{karakas07b}
and most of the new models here (see below) include an 
approximate treatment for the molecular opacities
(in particular CN, CO, H$_{2}$O, TiO) using the
formulations from \citet{bessell89} and corrected by
\citet{chiosi93}. These fits do include some compositional
dependence, but do not account for large variations in
C/O or nitrogen. The second prescription we use
is to include the low-temperature opacity tables from
\citet{ferguson05} in place of the \citet{bessell89}
fits.

The new AGB structure models are presented in 
Section~\ref{agbmodels}. Some of these models
have \citet[][hereafter VW93]{vw93} mass loss 
on the AGB, as do the models presented in 
\citet{karakas07b}.  We also compute new
intermediate-mass $Z= 0.0001$ AGB models 
using the Reimer's mass-loss formula 
\citep[][hereafter R75]{reimers75}, and use them 
in place of the VW93 models presented in 
\citet{karakas07b}. The main reason for this is to use 
a higher mass loss rate than given by VW93 at this
metallicity, which would allow the 6$\Msun$, 
$Z = 0.0001$ model to be evolved to very low envelope
mass. The VW93 6$\Msun$ model presented
in \citet{karakas07b} had $\sim 100$ TPs with very 
little reduction in envelope mass. The VW93 models are
compared to the R75 models in Section~\ref{agbmodels}.

We scale the R75 mass-loss formula to the metallicity 
of the model according to
\begin{equation}
  \frac{dM}{dt} = \sqrt{\left(Z/Z_{\odot}\right)} 
  \times \dot{M}_{\rm R} \label{mdot},
\end{equation} 
where ${dM}/{dt}$ is in $\Msun \, {\rm yr}^{-1}$,
$Z$ the global metallicity of the stellar model, 
and $Z_{\odot}$ the solar metallicity (0.02), and
\begin{equation}
   \dot{M}_{\rm R} =  \times 10^{-13} \eta_{\rm R}
\frac{(R/\Rsun)(L/\Lsun)} {(M/\Msun)}, \label{reimers}
\end{equation}
here $R$, $L$ and $M$ are the radius, luminosity and mass 
of the star (in solar units), respectively. We note
that \citet{decressin04} apply a similar $Z$ scaling
to their low-metallicity, intermediate-mass AGB models.
The free parameter $\eta_{\rm R}$ used for each model
is provided in Section~\ref{agbmodels}.
Equation~\ref{reimers} was used on the the first 
giant branch with $\eta_{\rm R} = 0.4$ in all stellar 
models.

The new stellar structure models are mostly
computed with the same version of the Monash stellar 
structure code used in \citet{karakas07b}.
The intermediate-mass $Z= 0.0001$ AGB models of 4, 4.5,
5, 5.5, and 6$\Msun$ are computed using an updated version
of the structure code that includes \citet{ferguson05}
low-temperature opacities. To check for consistency,
we compute a 4$\Msun$, $Z=0.0001$ model using both versions 
of the code. The results of the models are presented in 
Section~\ref{agbmodels}.

The technique used in the post-processing nucleosynthesis 
computations have been described in detail elsewhere,
see for example \citet{karakas09}. Here we summarize 
information pertinent to the current models. 
We assume a network of 77 species from
hydrogen to sulphur, along with a small group of iron-peak 
elements. Hence this study is limited to the nucleosynthesis
of light elements. Yields of heavy elements produced
by the $s$ process using the same codes are now 
becoming available, see \citet{karakas09} and \citet{church09}.
Most of the 589 reaction rates are from the JINA REACLIB 
database \citep{sakharuk06}\footnote{http://groups.nscl.msu.edu/jina/reaclib/db/}.
The details of the reaction rates used are described 
in Section~\ref{rates}.
We assume $Z=0.02$ for the solar composition and take the 
initial abundances from \citet{anders89}. For all other models
we assumed a scaled solar initial composition for all 
species.

\subsection{Updated reaction rates} \label{rates}

The reaction rate network used in the post-processing
calculations
is the recommended JINA REACLIB library \citep{sakharuk06} 
with the following changes.  We included the recommended 
rate for the \iso{18}F($\alpha$,$p$)\iso{21}Ne reaction from 
\citet{LeeThesis} and used in \citet{karakas08}. 
We used the \iso{22}Ne($\alpha,n$)\iso{25}Mg and
\iso{22}Ne($\alpha,\gamma$)\iso{26}Mg rates 
from \citet{karakas06a}, and the \iso{19}F($\alpha,p$)\iso{22}Ne
rate from \citet{ugalde08}. 

Our current reaction rates 
differ significantly to the rates
used in \citet{karakas07b} for the following reasons.
First, the NeNa and MgAl chain proton-capture rates
have been updated. The most important changes are the 
\iso{21}Ne($p, \gamma$)\iso{22}Na rate from \citet{iliadis01},
the \iso{22}Ne($p, \gamma$)\iso{23}Na rate from
\citet{hale02}, and the \iso{23}Na($p,\gamma$)\iso{24}Mg and 
\iso{23}Na($p,\alpha$)\iso{20}Ne rates
from \citet{hale04}.  The \iso{24}Mg($p, \gamma$)\iso{25}Al, 
the \iso{25}Mg($p, \gamma$)\iso{26}Al$^{g,m}$ (where
\iso{26}Al$^{g,m}$ are the ground and meta-stable state of
\iso{26}Al respectively), the \iso{26}Mg($p, \gamma$)\iso{27}Al
and the \iso{27}Al($p, \gamma$)\iso{28}Si and 
\iso{27}Al($p, \alpha$)\iso{24}Mg reaction rates are taken
from \citet{iliadis01}. Other important updates include
the most recent evaluation of the \iso{14}N($p,\gamma$)\iso{15}O
CNO rate from \citet{bemmerer06}, 
and the revised triple-$\alpha$ rate from \citet{fynbo05}.
Many of the rates are now from the NACRE compilation 
\citep{angulo99} including the following CNO reactions: 
\iso{12}C($p,\gamma$)\iso{13}N, \iso{13}C($p,\gamma$)\iso{14}N,
 \iso{16}O($p,\gamma$)\iso{17}F, \iso{18}O($p,\gamma$)\iso{19}F 
and \iso{18}O($p,\alpha$)\iso{15}N rates. The
\iso{17}O($p,\gamma$)\iso{18}F and 
\iso{17}O($p,\alpha$)\iso{14}N rates are from  
the more recent \citet{chafa07}.

The neutron source reactions have also been updated
including the
\iso{13}C($\alpha,n$)\iso{16}O rate which is taken from NACRE, 
whereas previously we used the \citet{denker95} rate. 
Previously we used the \citet{kaeppeler94} rates for the
\iso{22}Ne($\alpha,n$)\iso{25}Mg and 
\iso{22}Ne($\alpha,\gamma$)\iso{26}Mg reactions, 
now we use the rates from \citet{karakas06a}.

Some of the important $\beta$-decay rates have also been
updated or corrected. For example, the REACLIB database
used in \citet{karakas07b} had an incorrect decay rate 
for \iso{60}Fe of 0.22~Myr; this has been corrected to
1.5~Myr in the current JINA REACLIB library.

\subsection{The inclusion of a partial mixing zone} \label{sectionpmz}

In AGB stars there are two important neutron producing
reactions. The \iso{22}Ne($\alpha, n$)\iso{25}Mg reaction
operates during the convective thermal pulses when
$T \gtrsim 300 \times 10^{6}$K. This has 
been suggested to be  the dominant neutron source in
intermediate-mass AGB stars, whereas these temperatures 
are reached only in the last few TPs of lower mass
($M \lesssim 3\Msun$) stars.  The other potential source 
of neutrons in AGB stars is the 
\iso{13}C($\alpha,n$)\iso{16}O reaction, which operates at 
lower temperatures ($T \gtrsim 90 \times 10^{6}$) than the 
\iso{22}Ne source.   Observational and theoretical evidence 
suggests this is the dominant neutron source in low-mass
AGB stars \citep{smith87,gallino98}.

To operate efficiently the \iso{13}C($\alpha,n$)\iso{16}O 
reaction requires more \iso{13}C than is left over from CN 
cycling in the H-shell. 
Hence some mechanism to mix protons from the H-rich envelope 
into the intershell is needed to produce the extra \iso{13}C.
In our models, protons are mixed into the intershell 
region by artificially adding a partial mixing zone (PMZ) 
at the deepest extent of each TDU. 
These protons are captured by the abundant \iso{12}C 
to form \iso{13}C and \iso{14}N, resulting in the formation 
of a \iso{13}C pocket. In the \iso{13}C pocket, neutrons 
are liberated by the reaction
\iso{13}C($\alpha,n$)\iso{16}O during the interpulse 
period \citep{straniero95}.

The timescale for neutron production and the 
neutron source determine the resulting
$s$-process element distribution. The details of how the 
\iso{13}C pocket forms and its extent in mass in the 
He-intershell are still unknown, although various mechanisms 
have been proposed including convective overshoot, rotation, 
and gravity waves; see \citet{herwig05} for a discussion of 
the relative merits of each mechanism. In this study
we are concerned with the effect of a PMZ and the 
\iso{13}C($\alpha,n$)\iso{16}O on light-element 
nucleosynthesis. The neutrons from this reaction
are important for the production of e.g., \iso{19}F
\citep{forestini92,lugaro04}. The method we use to 
include a PMZ has been described in \citet{lugaro04}, 
and is similar to that used by \citet{goriely00}. 
We include a PMZ of constant mass
of 0.002$\Msun$ at the deepest extent of each third 
dredge-up episode for the 3$\Msun$, $Z = 0.02$, 0.008,
and 0.004 models, and for the 2$\Msun$, $Z=0.0001$
model.  

Here we only examine the effect of the PMZ on low-mass
AGB models. This is because we add the partially-mixed zone 
in the post-processing step, so any potential feedback 
on the structure of the star is 
ignored. In intermediate-mass models of 
low-metallicity ($M \gtrsim 3\Msun$ at $Z \le 10^{-4}$)
the temperature at the base of the convective envelope 
during dredge-up may become hot enough for proton-captures
\citep{goriely04,herwig04a}. 
The energy produced by these {\em hot dredge-ups} 
may effect the structure of the star, by increasing the 
depth of dredge-up \citep{herwig04a}, or by terminating
the AGB altogether \citep{woodward08}. This situation 
could arise if the ingestion of protons leads to an 
hydrogen flame that produces enough energy to eject
the envelope \citep{woodward08}. Consequences of proton 
ingestion on the nucleosynthesis is largely unknown 
but could include the 
inhibition of formation of the \iso{13}C pocket 
\citep{goriely04}.

The models we consider in this study do not fall into
the region of the mass--$Z$-plane where hot dredge-ups are 
predicted to occur \citep[see Fig.~4 from][]{goriely04}. 
The 3$\Msun$, $Z = 0.004$ model is the only candidate
for hot dredge-ups and in this model the peak temperature 
is $\lesssim 30 \times 10^{6}$~K during a few dredge-up 
episodes, but accompanied by low densities 
($\rho \lesssim 1$g/cm$^{3}$). It is unlikely
that proton captures occur during the TDU under these
conditions, although further study into the inclusion of a
PMZ on the structure of the 3$\Msun$, $Z = 0.004$ model 
is required. Note that we do not include a PMZ into the 
3$\Msun$, $Z=0.0001$ model, as this would suffer
hot dredge-ups, where the peak temperature at the base 
of the envelope during dredge-up is $\approx 60\times 10^{6}$~K.

Even if proton ingestion is unlikely to occur in our 
low-mass AGB models, these studies do point out the
inconsistency of adding a PMZ into the post-processing 
calculation. For this reason the results presented in 
Section~\ref{pmzyields} should
be treated with some caution. The trends described
in this paper are likely to be qualitatively accurate 
but the actual extent of the effect of a PMZ may be
different to that described here. Further detailed
studies into the effects of a partial mixing zone on
AGB nucleosynthesis is required, and will be the 
subject of a future investigation.

\section{The AGB models} \label{agbmodels}

\begin{table}
\begin{center}
\caption{Grids of stellar masses for each $Z$, noting if the models 
experience the core He-flash (CHe), the third dredge-up (TDU), and 
hot bottom burning (HBB). The second line lists the mass-loss law 
used on the AGB, including the parameter $\eta$ if the Reimers 
mass-loss formula was used.}\label{grid}
\begin{tabular}{cllll}
\hline Mass &  $Z = 0.02$ & $Z = 0.008$ & $Z = 0.004$ & $Z=10^{-4}$ \\
\hline 
\hline
1.0 & CHe & CHe & CHe & CHe,TDU \\
    & VW93 & VW93 & VW93 & VW93 \\
1.25 & CHe & CHe & CHe & CHe,TDU \\
    & VW93 & VW93 & VW93 & VW93 \\
1.5  & CHe & CHe & CHe,TDU & CHe,TDU \\
    & VW93 & VW93 & VW93 & VW93 \\
1.75 & CHe & CHe,TDU & CHe,TDU & CHe,TDU \\
    & VW93 & VW93 & VW93 & VW93 \\
1.9  & CHe & CHe,TDU & CHe,TDU & CHe,TDU  \\
    & VW93 & VW93 & VW93 & VW93 \\
2.0(2.1) & CHe &  CHe,TDU &  CHe,TDU & TDU \\
    & VW93 & VW93 & VW93 & VW93 \\
2.25 & CHe,TDU & TDU & TDU & TDU \\
    & VW93 & VW93 & VW93 & VW93 \\
2.5  & TDU & TDU & TDU & TDU \\
    & VW93 & VW93 & VW93 & VW93 \\
3.0  & TDU & TDU & TDU & TDU,HBB \\
     & VW93 & VW93     & VW93 & R75,$\eta=5$ \\
3.5  & TDU & TDU & TDU & TDU,HBB \\
     & VW93 & VW93     &   VW93 & R75,$\eta=7$ \\
4.0  & TDU & TDU,HBB & TDU,HBB & TDU, HBB \\
     & VW93 & VW93   &   VW93  & R75,$\eta=7$ \\
4.5  & TDU,HBB & TDU,HBB & TDU,HBB & TDU, HBB \\
     & VW93 &  VW93  &  VW93   & R75,$\eta=10$ \\
5.0  & TDU,HBB & TDUHBB & TDU,HBB & TDU,HBB \\
     & VW93  &  VW93  &  VW93   & R75,$\eta=10$ \\
5.5  &  TDU,HBB & TDUHBB & TDU,HBB & TDU,HBB \\
     & VW93 &  VW93   &  VW93   & R75,$\eta=10$ \\
6.0  & TDU,HBB & TDU,HBB & TDU,HBB & TDU,HBB \\
     & VW93 & VW93  &  VW93  & R75,$\eta=10$ \\
6.5  & TDU,HBB & --  & --  & --  \\
     & VW93 & -- & -- & -- \\
\hline
\hline
\end{tabular}
\medskip\\
\end{center}
\end{table}

The grids of AGB models used to compute the updated
yields are provided in Table~\ref{grid}. In this
table it is noted if the model experienced the core 
He-flash, TDU and/or HBB, and the mass loss 
formulation employed on the AGB. Most of the models 
listed in Table~\ref{grid} have
been previously described in detail 
\citep[e.g.,][]{karakas03a,karakas07b,karakas09}.
Here we restrict the discussion to the new AGB
models, which are the 4.5 and 5.5$\Msun$, $Z=0.02, 
0.008$ and 0.004, the 2.1$\Msun$, $Z = 0.004$, 
and the 1.0, 1.5, 1.9$\Msun$, $Z=0.0001$ models with VW93
mass loss. The new intermediate-mass $Z= 0.0001$ 
AGB models of 3.0, 3.5, 4.0, 4.5, 5.0, 5.5, and 6$\Msun$ 
are computed using R75 as described 
in Section~\ref{numerics}. 
The parameter $\eta_{\rm R}$ used for each  model 
is listed in Table~\ref{grid}. These parameters
were chosen as follows. Little is known about mass
loss from low-metallicity AGB stars of 
intermediate mass. However, detailed evolution
and nucleosynthesis models of such stars
indicate that the total C$+$N$+$O content of the
envelope quickly reaches similar levels found in 
AGB stars of much higher metallicity owing to
a very efficient TDU \citep[see in particular, 
\S2.1 from][]{herwig04b}. Hence there is no good
reason to suspect that the mass loss from these
stars should be significantly lower than, say,
AGB stars of $Z=0.004$. For this reason, we choose
the $\eta_{\rm R}$ parameters such that the final
number of thermal pulses is similar to that found
in $Z = 0.004$ models of the same mass. We note
that these choices are arbitrary, and changes in
$\eta_{\rm R}$ will have a strong impact on the 
chemical yields.

In Table~\ref{newmodels} we present some 
structural details from the new models including
the initial mass and metallicity, the final core
and envelope mass ($M_{\rm core}$ and $M_{\rm env}$, 
respectively), the number of TPs 
computed, and the total amount of matter dredged up 
into the envelope during the TP-AGB phase 
($M_{\rm dred}^{\rm tot}$). All masses are in
solar units.
There are two entries for the 4$\Msun$, $Z=0.0001$
case. The first line shows the results computed
with the code that includes \citet{ferguson05} 
low-temperature opacities, the second with the 
older version used to compute the models listed 
above that in Table~\ref{newmodels}.  The most 
important parameters, from a nucleosynthetic point 
of view, are the total number of thermal pulses as 
this determines how much material from the He-shell 
is mixed into the envelope, and the efficiency
of  HBB. These numbers are similar, although 
the model computed with the older code ended
with a higher final envelope mass and could, in
principle, experience at least one more TP. 
The effect of the (scaled solar composition) 
\citet{ferguson05} low-temperature opacities
on the evolution is small at such low metallicities,
and is the main reason for the consistency between
the two 4$\Msun$ models. The use of carbon-rich low
temperature opacities may have a much larger
impact, owing to the fact that all of the $Z=0.0001$
models, even the 1$\Msun$ and 6$\Msun$ models, 
become carbon rich with a final C/O $> 1$.

We provide an electronic on-line table with details 
of each new stellar model as a function of TP number;
(Table~A1).
This data are similar to the tables presented in 
\citet{karakas07b} but with the inclusion of the radius, 
bolometric luminosity, and effective temperature.
These tables include,
for each TP, the core mass ($M{\rm core}$), the maximum 
extent of the flash-driven convective region ($M_{\rm csh}$), 
the duration of flash-driven convection ($t_{\rm csh}$), the 
amount of matter dredged into the envelope after 
that pulse (Ddredge), the third dredge-up 
efficiency parameter ($\lambda$), and $L{\rm dup}$ defined 
by \citet{goriely00} to be $\Delta M_{\rm dredge}/M_{\rm csh}$.
The TDU efficiency parameter, $\lambda$, is usually defined 
according to $\lambda = \Delta M_{\rm dredge}/\Delta M_{\rm h}$, 
where $\Delta M_{\rm h}$  is the amount by which the core mass 
has grown between the present and previous TPs. 

Further, we provide the maximum temperature in the 
He-shell (THeshell), the maximum temperature 
at the base of the convective envelope during the 
previous interpulse period (Tbce), the maximum
temperature in the H-shell (THshell), the 
interpulse period, the total mass at the beginning 
of the TP, the maximum radiated luminosity during
the previous interpulse period (MaxL), the maximum
luminosity generated by the TP (MaxLHe), the maximum
radius during the previous interpulse period (maxR),
the bolometric luminosity ($M{\rm bol}$), and
finally, the effective temperature ($T{\rm eff}$).
All units are in solar units with the exception of 
temperatures, that are in kelvin, and all times 
which are in years.

\begin{table}
\begin{center}
\caption{Details of the new AGB models. 
}\label{newmodels}
\begin{tabular}{clllll}
\hline Mass &  $Z$ & $M_{\rm core}$ &$M_{\rm env}$ & No. TP & 
$M_{\rm dred}^{\rm tot}$ \\
\hline 
\hline
4.5 & 0.02 & 0.853 & 0.692 & 21 & 0.053 \\
5.5 & 0.02 & 0.900 & 0.800 & 36 & 0.075 \\
4.5 & 0.008 & 0.861 & 0.670 & 38 & 0.141 \\
5.5 & 0.008 & 0.907 & 0.864 & 56 & 0.143 \\
2.1 & 0.004 & 0.650 & 0.090 & 20 & 0.058 \\
4.5 & 0.004 & 0.873 & 1.160 & 50 & 0.177 \\
5.5 & 0.004 & 0.932 & 0.484 & 71 & 0.154 \\
1.0 & 0.0001 & 0.726 & 8.5($-3$) & 26 & 2.6$(-3)$ \\
1.5 & 0.0001 & 0.662 & 0.022 & 18 & 0.06 \\
1.9 & 0.0001 & 0.682 & 0.029 & 24 & 0.170 \\ 
3.0 & 0.0001 & 0.812 & 0.120 & 20 & 0.106 \\
3.5 & 0.0001 & 0.854 & 0.136 & 27 & 0.103 \\
4.0 & 0.0001 & 0.872 & 0.133 & 37 & 0.128 \\
    &        & 0.876 & 0.477 & 36 & 0.115 \\
4.5 & 0.0001 & 0.898 & 0.085 & 41 & 0.109 \\
5.0 & 0.0001 & 0.929 & 0.123 & 56 & 0.122 \\
5.5 & 0.0001 & 0.966 & 0.085 & 77 & 0.124 \\
6.0 & 0.0001 & 1.008 & 0.183 & 109 & 0.127 \\
\hline
\hline
\end{tabular}
\medskip\\
\end{center}
\end{table}

We briefly comment on the 1$\Msun$, $Z=0.0001$
model. This model experienced 26 TPs
and had a final core mass of 0.72$\Msun$,
higher than found in metal-rich models
of the same initial mass (e.g., 0.56$\Msun$ 
at $Z=0.02$). The 1$\Msun$, $Z=0.0001$ experienced
shallow TDU after the 5$^{\rm th}$ and 6$^{\rm th}$ 
TPs (the core mass was $\approx 0.56\Msun$ at the
6$^{\rm th}$ TP). The small amount of TDU was 
enough to make the star carbon rich, 
where C/O $>1$. The addition of primary \iso{12}C
into the envelope had a dramatic effect on the 
evolution of the star. This is because the extra 
\iso{12}C changed the rate of nuclear burning in 
the H-shell. The additional energy caused a 
significant shortening of the interpulse period
from  $\sim$440,000~years at the 6$^{\rm th}$  TP
to 240,000~years, followed by a steady decrease
to 45,000~years, and there was no further TDU.
Similar behaviour was reported by \citet{stancliffe08} 
for the same mass and metallicity. 

For the remainder of this section we compare the 
R75 models to the VW93 models from \citet{karakas07b}. 
We will use the 3 and 6$\Msun$, $Z=0.0001$ models 
as examples.  The 3$\Msun$ model with
VW93 had 40 TP and dredged a total of 0.242$\Msun$
into the envelope. This model also experienced HBB
with a peak temperature of $71 \times 10^{6}$K.
The final core mass was 0.82$\Msun$.  In comparison,
the model with R75 mass loss had 20 TPs and 
dredged 0.106$\Msun$ into the envelope. This model 
only had mild HBB,
with a peak temperature of $\sim 40\times 10^{6}$K.
The final core mass was 0.81$\Msun$, probably the 
only similarity to the VW93 model. 

\begin{figure}
\begin{center}
 \includegraphics[width=5cm,angle=270]{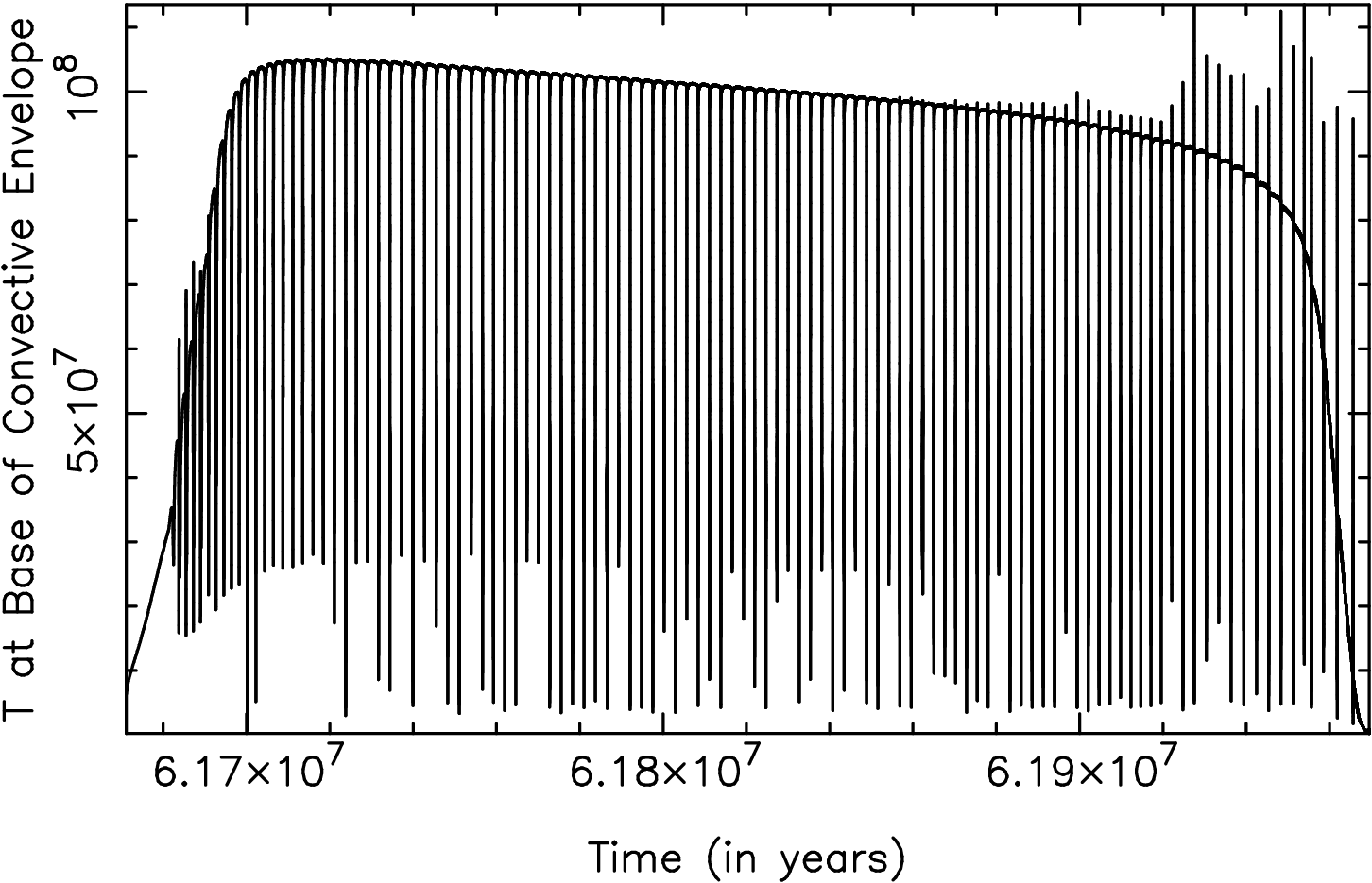}
 \includegraphics[width=5cm,angle=270]{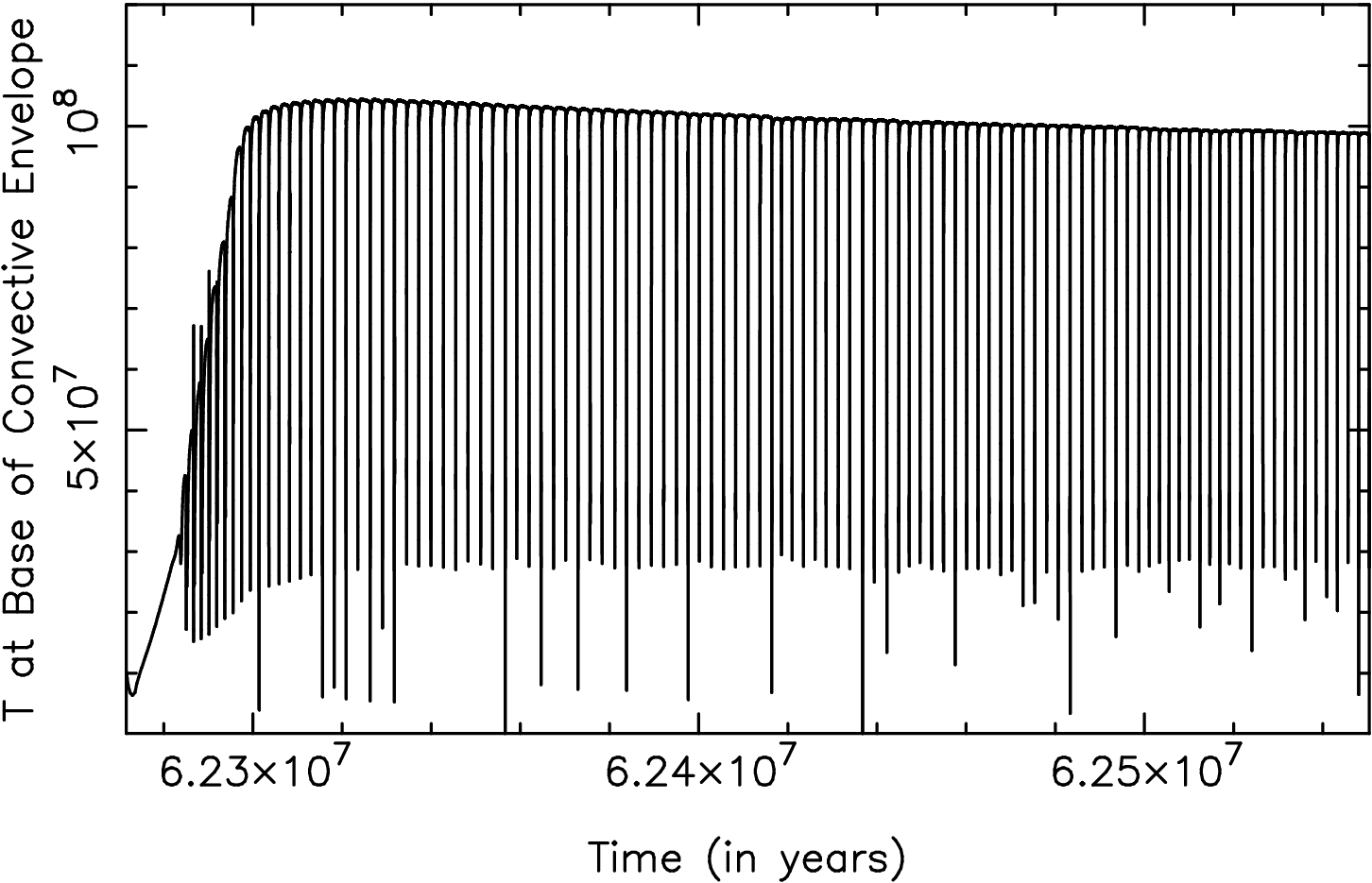}
 \caption{The temporal evolution during the TP-AGB
of the temperature at the base of the convective envelope 
for 6$\Msun$ $Z = 0.0001$ models with R75 (top panel) and 
VW93 (lower panel) mass loss.
\label{fig1}}
\end{center}
\end{figure}

In contrast, both 6$\Msun$, $Z = 0.0001$ models 
had about the same number of TPs: 106 in
the VW93 case compared to 109 in the case of 
the R75 model. The peak temperature at the base
of the envelope was also similar at 
$104 \times 10^{6}$K, along with  the total amount
of matter dredged into the envelope, (0.114$\Msun$ 
for the VW93 model compared to 0.127$\Msun$).
The real difference between the models is that
the R75 model was evolved to a small envelope
mass of only $0.13\Msun$, which led to the 
cessation of HBB. 
This can be seen in Figure~\ref{fig1} where
we illustrate the evolution of the temperature at
the base of the convective envelope during the 
TP-AGB for the R75 (top panel) and VW93 (lower 
panel) models. In the top panel the temperature 
drops dramatically to below $10 \times 10^{6}$K 
during the last few TPs. This is caused by the
rapid reduction in the envelope mass. In comparison 
the VW93 model (top panel) only loses 
$\sim 0.04\Msun$ during the AGB and we expect 
that this model may experience at least another
$\sim 100$~TPs, setting the yields presented in 
\citet{karakas07b} as lower limits.

\section{Stellar Yields} \label{yields}

To compute the yields we integrate the mass lost 
from the model star during the entire stellar lifetime 
according to
\begin{equation}
 M_{i} = \int_{0}^{\tau} \left[ X(i) - X_{0} (k)\right] 
\frac{d M}{dt} dt,
\label{yield}
\end{equation}
where $M_{i}$ is the yield of species $i$ (in solar masses),
$dM/dt$ is the current mass-loss rate, $X(i)$ and $X_{0} (i)$ 
refer to the current and initial mass fraction of species $i$, 
and $\tau$ is the total lifetime of the stellar model. The 
yield can be negative, in the case where the element is destroyed, 
and positive if it is produced.
We also present the total amount of $i$ (in $\Msun$) expelled into
the ISM, noting that this value is always positive.

The stellar yields for the models listed in Table~\ref{grid}
are available as electronic on-line tables. We provide 
one table for each metallicity (Tables~A2 to A5).
in the Appendix for an example.
Each table contains the following: 1) the nuclear species, $i$,
2) the atomic mass, {\it A}($i$), 3) the net stellar yield 
defined above, 4) the amount of species $i$ in the wind lost 
from the star, {\it mass}$(i)_{\rm lost}$, 5) the amount of 
$i$ that would have initially been present in the wind, 
{\it mass}$(i)_{0}$. The quantity {\it mass}$(i)_{0}$ is 
the mass expelled during the stellar lifetime multiplied 
by the initial mass fraction.
We next include 6) the average mass fraction of $i$ in 
the wind, $\langle X(i)\rangle$, 7) the initial mass 
fraction $X_{0}(i)$, and 8) the production factor $f$ 
defined by $\log_{10} [ \langle X(i)\rangle / X_{0}(i) ]$. 
All yields are in solar masses and abundances in mass
fraction.  These yields are
presented in the same format as in \citet{karakas07b}.
We note that there are two entries for \iso{26}Al: the first
shows the yields for the longer-lived ground state 
\iso{26}Al$^{\rm g}$ which has a half life of 
$7.17 \times 10^{5}$ years and is denoted by the symbol 
``al-6'' in the on-line tables; ``al$*6$'' refers to the 
short-lived meta-stable state. The yields of 
``al-6'' should be added to that of \iso{26}Mg for
chemical evolution studies.

In \citet{karakas07b} we discussed the inclusion of 
of synthetic TPs to account for TPs not computed in
detail.  This was done because 
some of the models had reasonably large envelope masses 
at the end of the computation and could, in principle,
experience further TPs and dredge-up. We do not repeat
that discussion here but we also do not include the 
contribution of these synthetic TPs for
the following reason. We would need to make an assumption
about the efficiency of the third dredge-up parameter
at small envelope mass, and this is a unknown. There
is evidence to suggest that $\lambda$ decreases
with decreasing envelope masses for low-mass 
\citep{straniero97,karakas02} and intermediate-mass 
AGB stars \citep{vw93}. 
The new intermediate-mass $Z=0.0001$ AGB models 
presented here were evolved to small envelope masses 
($\sim 0.1\Msun$ in some cases). These models do not
in general show any decrease in $\lambda$ with 
decreasing envelope mass. However, it should be 
pointed that low-metallicity models tend to experience 
more efficient mixing than metal-rich AGB stellar models
\citep[e.g.,][]{boothroyd88c,karakas02}.

One last difference between the yields presented
here and in \citet{karakas07b} is that we do not 
specifically provide yields for planetary nebulae (PNe). 
That is because column~6 of the on-line yield tables 
provides the average mass fraction of the wind. 
This value is weighted toward the composition of the 
envelope during the last few TPs, because that is when 
most of the mass is lost from the star. Hence this 
value is suitable for comparison to PNe abundances.
The models computed using the R75 mass-loss formula
have not had their mass loss weighted toward the final
few TPs, as they would have if we were to use the
VW93 prescription. However few, if any, PNe should exist 
that have evolved from intermediate-mass $Z=0.0001$
AGB stars.  The few PNe found in the Galactic Halo 
likely evolved from stars with initial masses closer 
to $\sim 1\Msun$.

\subsection{Comparison to previous work}

\begin{figure*}
\begin{center}
 \includegraphics[width=10cm,angle=270]{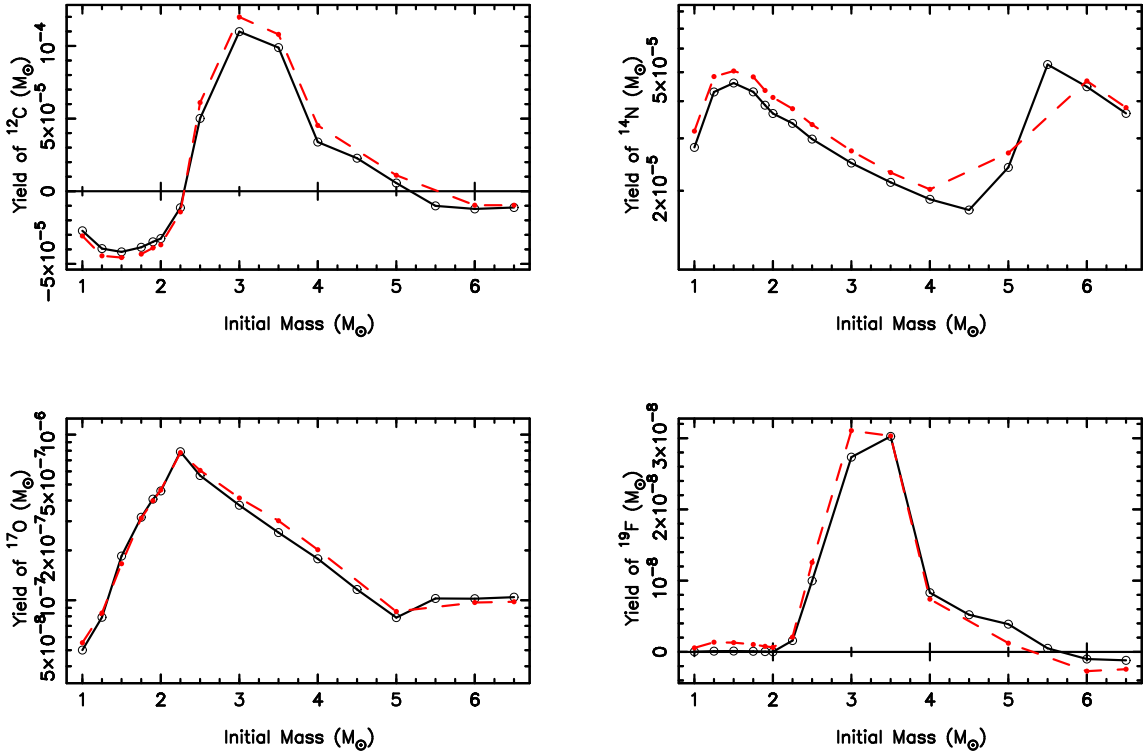}
 \caption{Weighted yields of \iso{12}C, \iso{14}N, 
\iso{17}O, and \iso{19}F as a function of the initial 
mass for the $Z=0.02$ models. The solid line and open
circles show results for the updated yields; the
dashed line and filled circles show results from 
\citet{karakas07b}.
\label{fig2}}
\end{center}
\end{figure*}

\begin{figure*}
\begin{center}
 \includegraphics[width=10cm,angle=270]{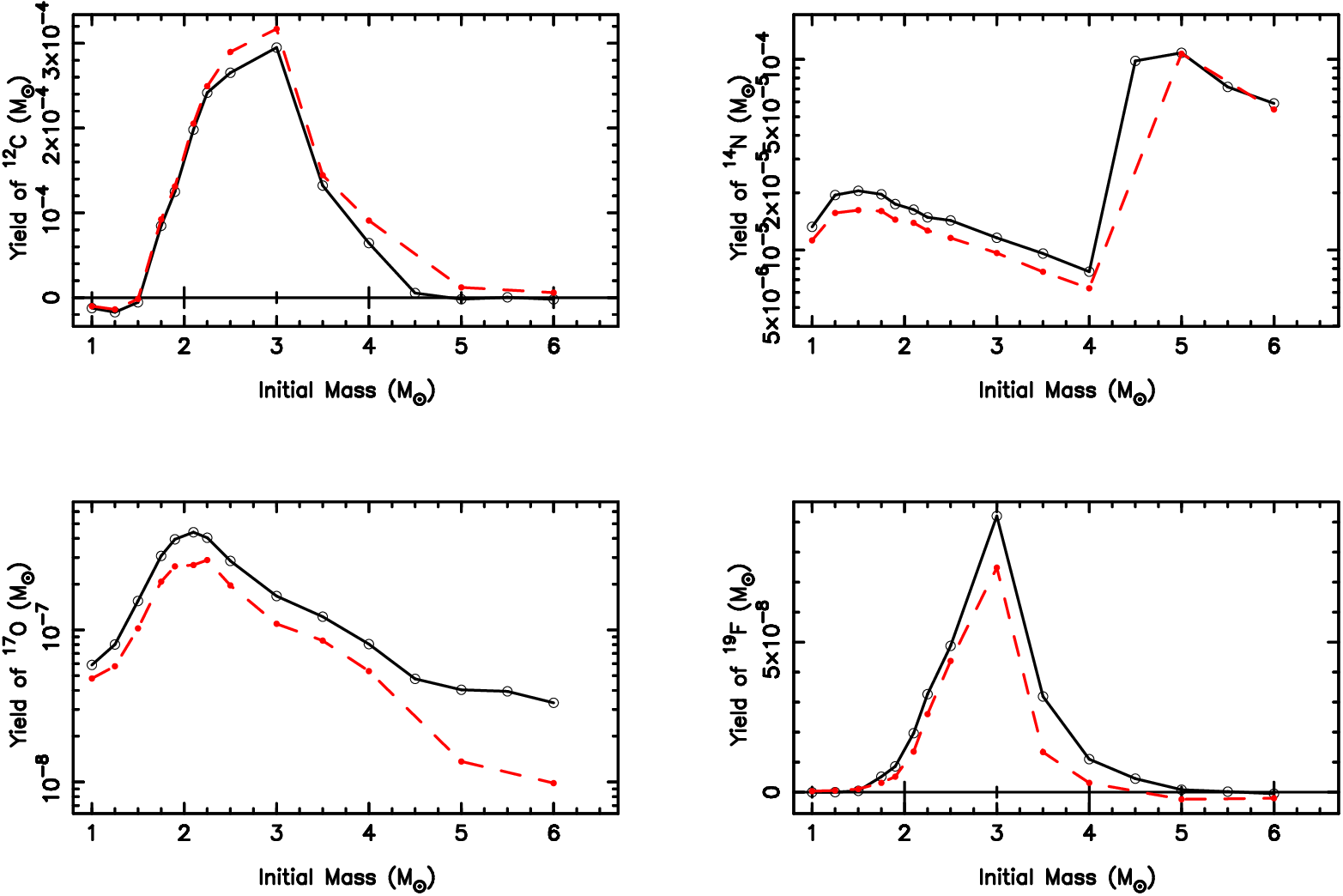}
 \caption{Weighted yields of \iso{12}C, \iso{14}N, 
\iso{17}O, and \iso{19}F as a function of the initial 
mass for the $Z=0.008$ models. Symbols are the same
as in Figure~\ref{fig2}.
\label{fig3}}
\end{center}
\end{figure*}

\begin{figure*}
\begin{center}
 \includegraphics[width=10cm,angle=270]{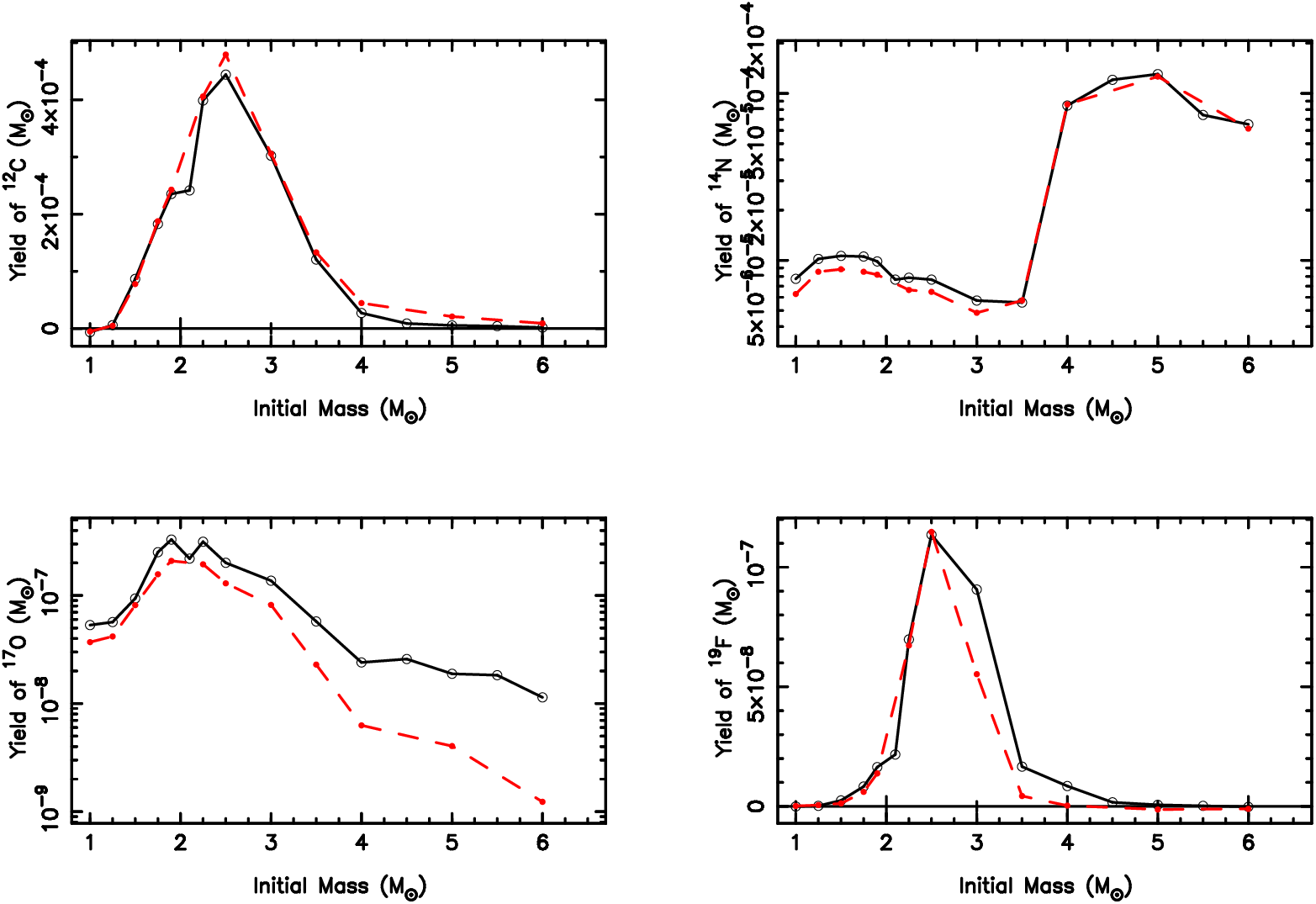}
 \caption{Weighted yields of \iso{12}C, \iso{14}N, 
\iso{17}O, and \iso{19}F as a function of the initial 
mass for the $Z=0.004$ models. Symbols are the same
as in Figure~\ref{fig2}.
\label{fig4}}
\end{center}
\end{figure*}

\begin{figure*}
\begin{center}
 \includegraphics[width=10cm,angle=270]{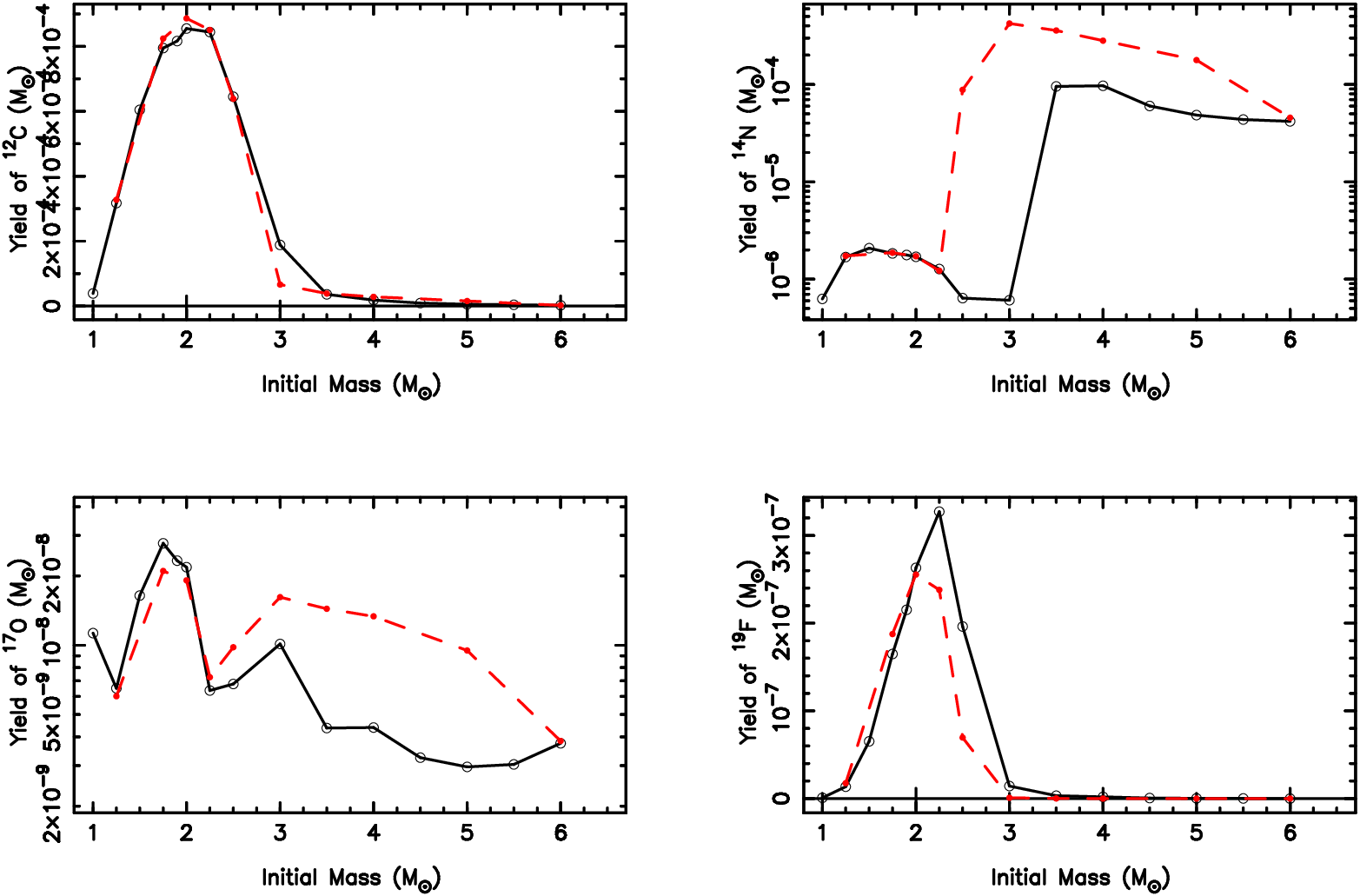}
 \caption{Weighted yields of \iso{12}C, \iso{14}N, 
\iso{17}O, and \iso{19}F as a function of the initial 
mass for the $Z=0.0001$ models. Symbols are the same
as in Figure~\ref{fig2}.
\label{fig5}}
\end{center}
\end{figure*}

In Figures~\ref{fig2} to~\ref{fig5} we show the weighted
yields of \iso{12}C, \iso{14}N, \iso{17}O, and \iso{19}F
from this study and from \citet{karakas07b}
for the $Z =0.02, 0.008, 0,004$, and $Z = 0.0001$ models,
respectively. In each plot, the yields have been weighted
by the initial mass function (IMF) from \citet{kroupa93}.
Note for this comparison that we are using the yields from 
\citet{karakas07b} with no contribution from synthetic TPs.
These isotopes were chosen to be representative of light
element nucleosynthesis in AGB stars of various mass.

At $Z = 0.02$, the new yields of \iso{12}C, \iso{17}O,
and \iso{19}F are reasonably consistent with the yields 
from \citet{karakas07b}.  We see small
reductions in the yield of \iso{12}C for $M > 2.5\Msun$,
and small increases in the yields of \iso{19}F for 
$M \ge 4\Msun$.  The yields of \iso{14}N
are systematically lower for masses less than 4$\Msun$,
whereas above this we can see the effect of a finer mass
grid on the shape of the yield as a function of stellar
mass. For the $Z=0.008$ and 0.004 models (Figures~\ref{fig3} 
and~\ref{fig4}) we see a similar trend as seen for 
$Z=0.02$, although the new yields of \iso{17}O and 
\iso{19}F are generally larger.
The \iso{12}C and \iso{14}N yields are remarkably 
consistent, indicating that the change of reaction rates
or initial abundances has not had a significant
impact on the yields of these important isotopes.

The new $Z=0.0001$ model yields are consistent with
the old yields for \iso{12}C and for \iso{19}F. The 
main difference for fluorine is that the peak production
is shifted to 2.5$\Msun$ from 2.25$\Msun$, as a 
consequence of using the lower 
\iso{19}F($\alpha,p$)\iso{22}Ne destruction rate 
from \citet{ugalde08}.
The main difference between the yields in Figure~\ref{fig5}
are for \iso{14}N and \iso{17}O. In both cases, the
new yields produce less of each isotope at a given
mass for $M \ge 2.5\Msun$.  The differences observed at
2.5$\Msun$ are surprising, since we are
using the same structure model. The changes in the
yields at this mass are a consequence of the new 
reaction rates (and in particular the new 
\iso{14}N($p,\gamma$)\iso{15}O and \iso{17}O $+p$
rates). However the difference for \iso{14}N
is most apparent at 3$\Msun$, where the variation
is a factor of $\sim 650$. If we assume that the
rates have introduced at least a factor of two variation,
then the change of structure model (using the R75
model instead of the VW93) is still by far the most 
significant factor. For \iso{17}O, the new 
intermediate-mass $Z=0.0001$ models result in 
smaller yields for $M \ge 3\Msun$, with the 
biggest variation at $4-5\Msun$.

\begin{figure*}
\begin{center}
 \includegraphics[width=5cm,angle=270]{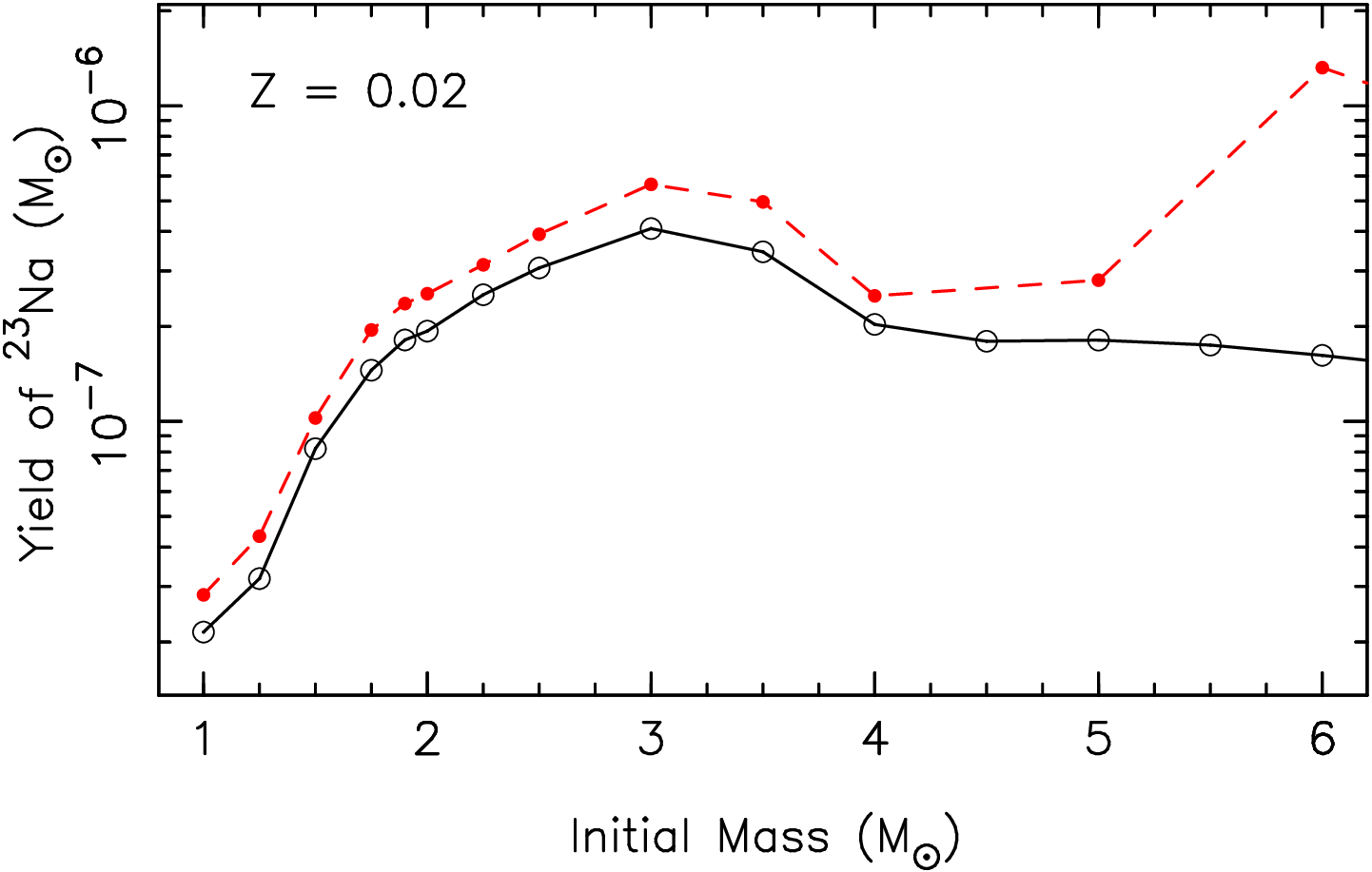}
 \includegraphics[width=5cm,angle=270]{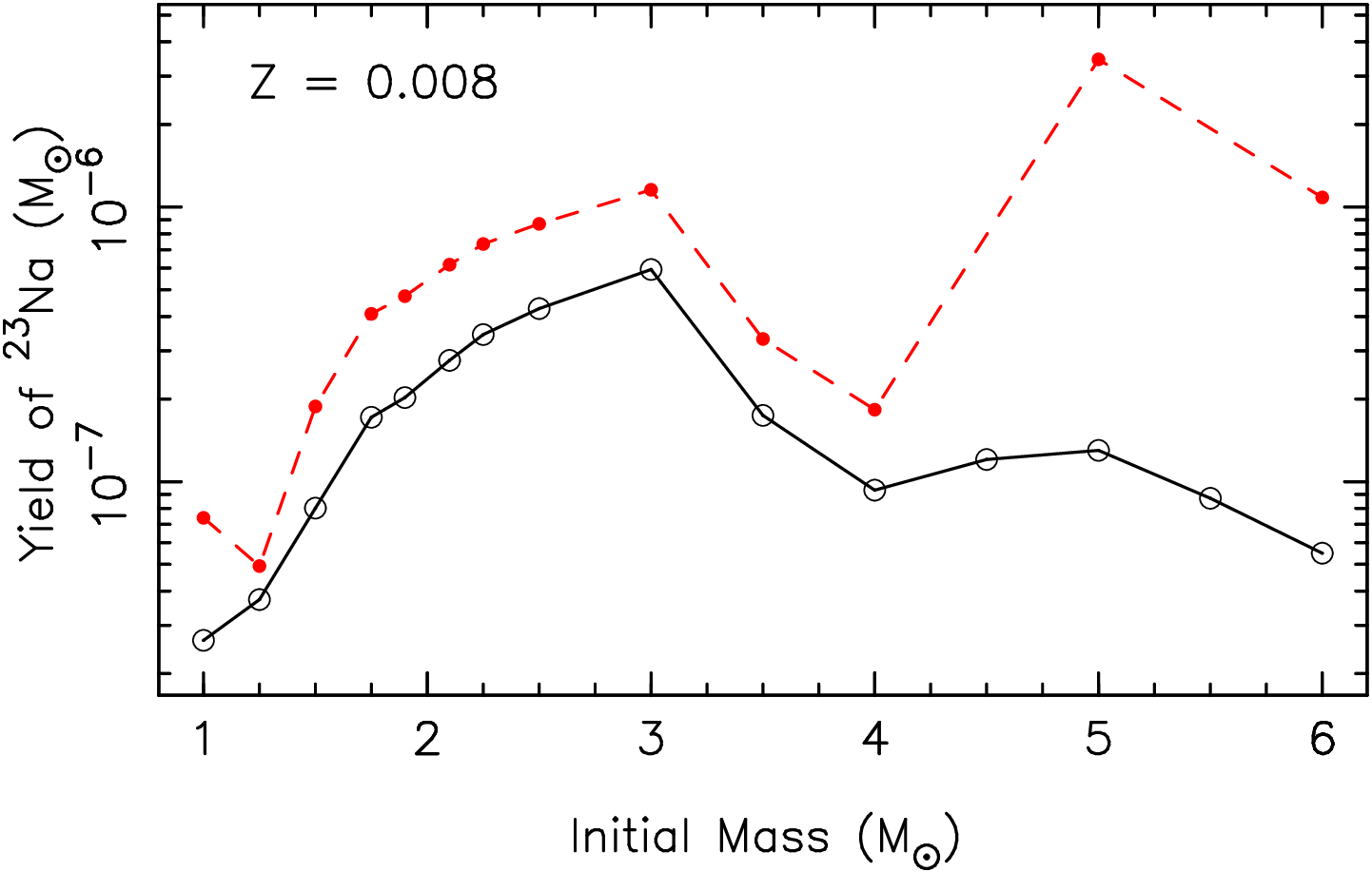}
 \includegraphics[width=5cm,angle=270]{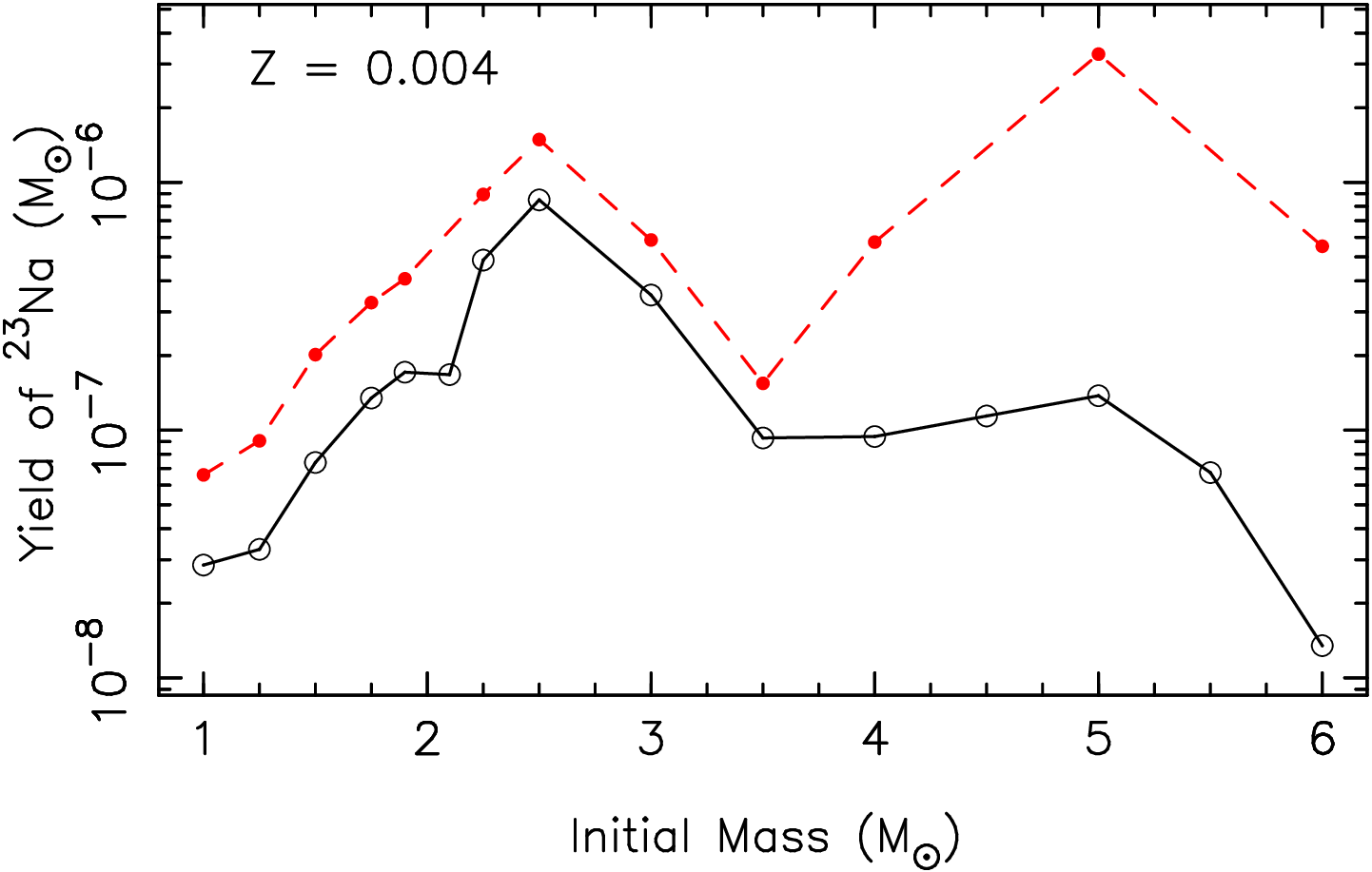}
 \includegraphics[width=5cm,angle=270]{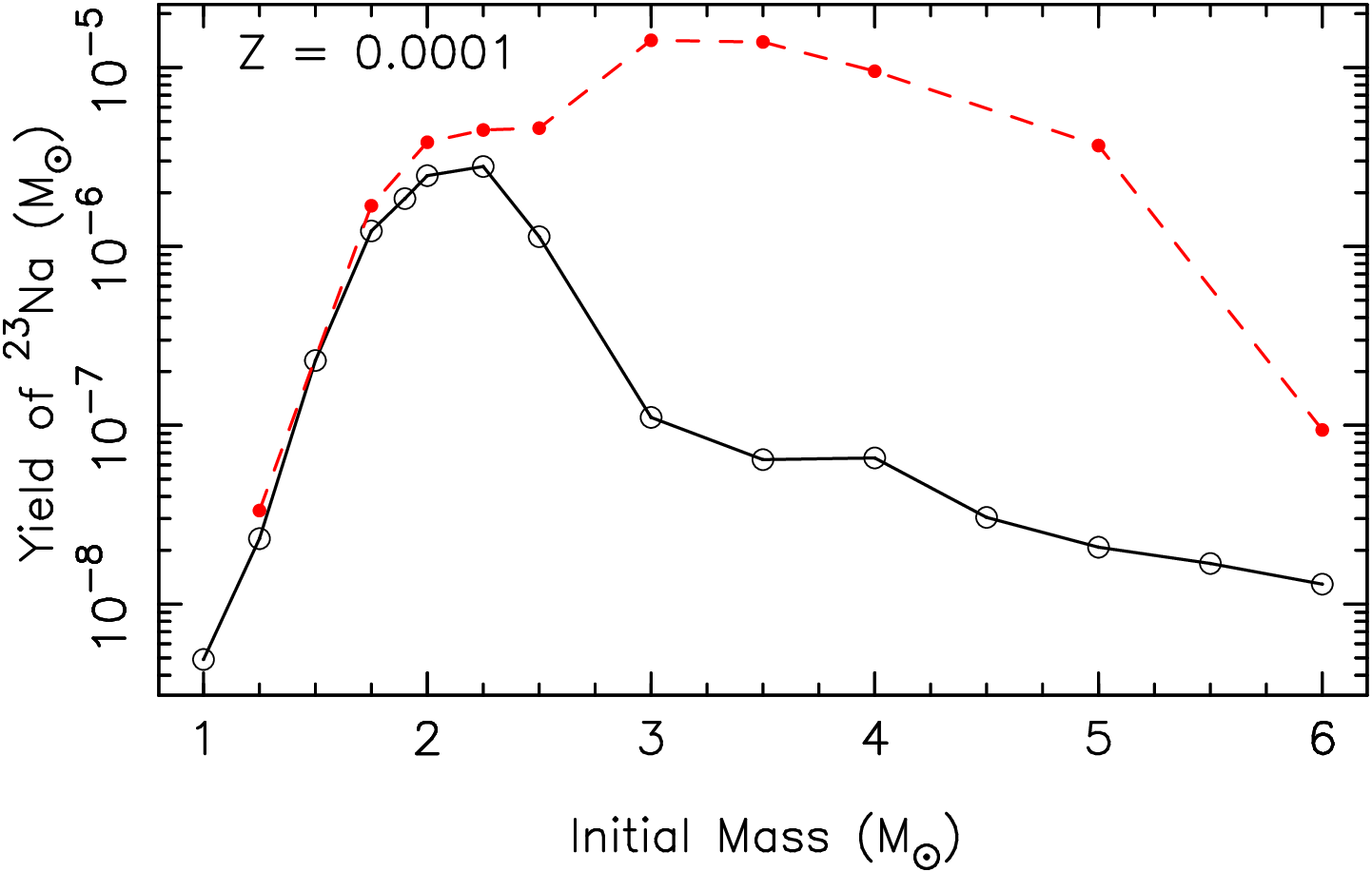}
 \caption{Weighted yields of \iso{23}Na for the
$Z = 0.02$, $Z= 0.008$, $Z = 0.004$, and $Z=0.0001$ 
models. Symbols are the same as in Figure~\ref{fig2}.
\label{fig6}}
\end{center}
\end{figure*}

In Figure~\ref{fig6} we show the weighted yields of
\iso{23}Na for the $Z=0.02$, 0.008, 0.004, and 0.0001 
models, again comparing the new yields to the yields 
from \citet{karakas07b}.  It is for this element that
we see the largest differences as a result of
using the new reaction rates.  The new yields
are systematically lower for all stellar models computed.
Table~\ref{na23} shows that the variation increases
with increasing mass, and at a given mass, with 
decreasing metallicity. That is, models with HBB show 
the largest impact of using the new \iso{23}Na($p,\gamma$)\iso{24}Mg 
and \iso{23}Na($p,\alpha$)\iso{20}Ne
reaction rates from \citet{hale04}, which are faster than
the \citet{eleid95} rates that we used previously.
This has resulted in higher rates of \iso{23}Na destruction 
and dramatically lower yields of sodium for $M \gtrsim 3\Msun$
for all metallicities. \citet{izzard07} found a 
similar result using both detailed and synthetic AGB 
models and a similar set of reaction rates.
\citet{ventura05b} discussed the variation of \iso{23}Na 
production as a result of uncertainties in reaction rates.
They compared the \citet{cf88} rates to the NACRE database,
and found similarly large differences as reported here.

\begin{table*}
\begin{center}
\caption{Yields (in $\Msun$) of \iso{23}Na from models with
$M \ge 3\Msun$. The old yield refers to the yields
from \citet{karakas07b}.
For the $Z= 0.0001$ models, we compare the sodium
yields from the R75 models presented here to the VW93 
models from \citet{karakas07b}. For the 6$\Msun$, $Z=0.0001$
there are two entries: the first shows the new and old
yields from the VW93 model, and the second line the new
yield for the R75 model. All yields
are expressed in the form $n(m)= n\times 10^{m}$.}
\label{na23}
\begin{tabular}{cclll} \hline \hline
Initial mass & $Z$ & New \iso{23}Na yield & Old \iso{23}Na yield &
Factor difference \\  \hline
3.0 & 0.02 & 5.0942($-$5) & 7.0317($-$5) & 1.38 \\
3.5 & 0.02 & 6.5188($-$5) & 9.3885($-$5) & 1.44 \\ 
4.0 & 0.02 & 5.5071($-$5) & 6.7809($-$5) & 1.23 \\ 
5.0 & 0.02 & 8.9715($-$5) & 1.3889($-$4) & 1.55 \\ 
6.0 & 0.02 & 1.3131($-$4) & 1.0715($-$3) & 8.16 \\
6.5 & 0.02 & 1.4863($-$4) & 1.0026($-$3) & 6.75 \\ \hline 
3.0 & 0.008 & 7.3897($-$5) & 1.4416($-$4) & 1.95 \\ 
3.5 & 0.008 & 3.2963($-$5) & 6.2545($-$5) & 1.90 \\ 
4.0 & 0.008 & 2.5255($-$5) & 4.9604($-$5) & 1.96 \\
5.0 & 0.008 & 6.4444($-$5) & 1.7090($-$3) & 26.5 \\   
6.0 & 0.008 & 4.4473($-$5) & 8.7833($-$4) & 19.7 \\ \hline
3.0 & 0.004 & 4.3768($-$5) & 7.2967($-$5) & 1.67 \\
3.5 & 0.004 & 1.7585($-$5) & 2.9209($-$5) & 1.66 \\ 
4.0 & 0.004 & 2.5553($-$5) & 1.5553($-$4) & 6.09 \\ 
5.0 & 0.004 & 6.8179($-$5) & 1.6316($-$3) & 23.9 \\ 
6.0 & 0.004 & 1.0933($-$5) & 4.4738($-$4) & 40.9 \\ \hline 
3.0 & 0.0001 & 1.3794($-$5) & 1.7719($-$3) & 128 \\
3.5 & 0.0001 & 1.2139($-$5) & 2.6254($-$3) & 216 \\
4.0 & 0.0001 & 1.7816($-$5) & 2.5865($-$3) & 145 \\
5.0 & 0.0001 & 1.0310($-$5) & 1.8136($-$3) & 176 \\ 
6.0 & 0.0001 & 7.0570($-$6) & 7.6308($-$5) & 10.8 \\ 
6.0 & 0.0001 & 1.0482($-$5) & --           & -- \\
\hline \hline
\end{tabular}
\medskip\\
\end{center}
\end{table*}

\subsection{The effect of a partial mixing zone} \label{pmzyields}

In this section we consider  the effect of a partially
mixed zone inserted at the deepest extent of each 
third dredge-up
episode on the stellar yields. In Table~\ref{pmz}
we show the yields for selected isotopes from the 3$\Msun$
models.  The yields for all species for each model 
with a PMZ is included as an on-line data table (Table~A6).
The effect of a PMZ on the nucleosynthesis 
of a 2$\Msun$, model of $Z= 0.0001$ is discussed in 
Section~\ref{otherauthors} in comparison to results 
from \citet{cristallo09}. In each model, a PMZ of 
0.002$\Msun$ has been inserted.  
Note that in the computations the same initial abundances
and reaction rate network are used. The only variation
is the inclusion of a partially mixed
zone at the deepest extent of each dredge-up episode.

The extra \iso{14}N in the intershell results in 
higher yields of \iso{22}Ne, and hence also \iso{23}Na.
The production of \iso{23}Na occurs via the
\iso{22}Ne($n,\gamma$)\iso{23}Ne($\beta^{-1}$)\iso{23}Na
sequence of reactions \citep{goriely00,herwig04b}, and 
is the main explanation for the increase in this element 
in Table~\ref{pmz}. 
The neutrons from the \iso{13}C($\alpha,n$)\iso{16}O
reaction are also responsible for enrichments
in other isotopes, including \iso{19}F, 
\iso{21}Ne, \iso{25}Mg, \iso{26}Mg, \iso{26}Al, 
\iso{27}Al, \iso{29}Si, \iso{30}Si, \iso{31}P, and
\iso{60}Fe.

The isotopes of \iso{31}P and \iso{60}Fe are only produced by 
neutron capture in AGB stars. The production of the 
radioactive \iso{60}Fe requires high neutron densities 
to overcome the branching at \iso{59}Fe. Hence \iso{60}Fe
production is primarily via the 
\iso{22}Ne($\alpha,n$)\iso{25}Mg source during TPs. 
Hence the 3$\Msun$, $Z = 0.02$ model without a PMZ produces 
very little \iso{31}P or \iso{60}Fe, as it is not hot enough 
to sustain the \iso{22}Ne($\alpha,n$)\iso{25}Mg source 
during TPs, and without a PMZ, does not have enough 
\iso{13}C present to produce any neutron capture elements.
The increase of \iso{60}Fe caused by the PMZ is the result
of the extra \iso{22}Ne in the intershell (from \iso{14}N)
causing a slightly higher activation of the
\iso{22}Ne($\alpha,n$)\iso{25}Mg reaction. 
The difference in the production of \iso{31}P and 
\iso{60}Fe is lessened in the 
lower metallicity models, mostly as a consequence of
partial activation of the \iso{22}Ne neutron source in
models without a PMZ.

\begin{table*}
\begin{center}
\caption{Yields for selected isotopes from the 3$\Msun$ 
models. The first line shows the yield
from the model with no PMZ, the second line the yield 
from the PMZ model, and the third line the percentage 
difference between the yields.}
\label{pmz}
\begin{tabular}{cllllllll} \hline \hline
$Z$    & \iso{19}F  & \iso{22}Ne & \iso{23}Na & \iso{25}Mg &
\iso{26}Mg  & \iso{30}Si & \iso{31}P &  \iso{60}Fe \\  \hline
0.02   & 3.4087($-6$) & 1.8092($-3$) & 5.0942($-5$) & 1.0767($-5$) & 
1.3568($-5$) & 7.7026($-7$) & 8.0023($-8$) & 9.5257($-9$) \\
       & 4.1292($-6$) & 3.4512($-3$) & 9.2811($-5$) & 2.1733($-5$) &
2.5835($-5$) & 3.1746($-6$) & 1.6082($-6$) & 2.2711($-8$) \\
       & 21.1\% & 90.8\% & 82.2\% & 102\% & 90.4\% &  312\% & 1910\% & 138\% \\ 
0.008  & 1.1481($-5$) & 4.2290($-3$) & 7.3897($-5$) & 6.7804($-5$) &  
7.3233($-5$) & 2.5349($-6$) & 2.7079($-7$) & 3.4270($-7$) \\
       & 1.3700($-5$) & 7.5472($-3$) & 1.6357($-4$) & 1.0954($-4$) &
1.3010($-4$) & 3.1024($-6$) & 2.1346($-6$) & 4.0355($-7$) \\
       & 19.3\% & 78.5\% & 121\% & 61.6\% & 77.6\% & 22.4\% & 688\% & 17.8\%  \\
0.004  & 1.1311($-5$) & 3.4789($-3$) & 4.3768($-5$) & 1.0715($-4$) & 
1.4453($-4$) & 2.9138($-6$) & 4.1899($-7$) & 1.4891($-6$) \\
       & 1.4083($-5$) & 6.9083($-3$) & 1.0657($-4$) & 1.8292($-4$) &
2.7583($-4$) & 3.8189($-6$) & 1.9773($-6$) & 1.7557($-6$) \\
       & 24.5\% & 98.6\% & 143\% & 70.7\% & 90.8\% & 31.1\% & 372\% & 18.0\%  \\
\hline \hline
\end{tabular}
\medskip\\
\end{center}
\end{table*}

\subsection{Comparison to other authors} \label{otherauthors}

In this section we compare the structure and yields 
from the 2$\Msun$, $Z=0.0001$ model to the comparative model
from \citet{cristallo09}. For the sake of the comparison 
we compute a 2$\Msun$, $Z = 0.0001$ model with a PMZ of 
0.002$\Msun$, which results in a \iso{13}C pocket that is 
$\sim 10$\% of the mass of the intershell.
This is because all the models in \citet{cristallo09} have
\iso{13}C pockets that formed as a consequence of the
algorithm used to find the convective border following
a TP. We assumed an exponentially-decaying proton profile
in our PMZ models and this results in a \iso{13}C pocket
that is similar in shape to the $\beta = 0.1$ case in 
\citet[][see the top panel of Fig.~4]{cristallo09}. 
The $\beta = 0.1$ case is their adopted value in the
full detailed nucleosynthesis computations.
An important difference is the mass extent of the pocket.
From the top panel of Fig.~4 in \citet{cristallo09},
we estimate\footnote{these mass estimates are smaller 
than the effective \iso{13}C pocket size defined 
in Cristallo et al. (2009). In either case
our \iso{13}C pockets are somewhat smaller.} 
that the extent of the \iso{13}C pocket 
(where the \iso{13}C abundance is larger than the
\iso{14}N abundance) to be $\approx 0.0015\Msun$, whereas
in our model we obtain  a value about 
1.5 times smaller, at $\approx 9\times 10^{-4}\Msun$. This mass 
difference has important implications for the nucleosynthesis
as discussed below.

The main difference between the models is the 
treatment of convective borders, and the use of carbon and
nitrogen-rich low-temperature opacities in the \citet{cristallo09}
model. The effect of C-rich opacities on AGB evolution 
was first discussed by \citet{marigo02}, 
who noted that once C/O $>1$ the star cools and mass loss
increases, shortening the TP-AGB lifetime. This has 
the effect of reducing the total number of TPs (Cristallo  
et al. find 15, we have 26), and the total amount
of mass dredged up into the envelope (our model dredges
up 30\% more He-shell material). 

\begin{table*}
\begin{center}
\caption{Yields for selected isotopes from the 2$\Msun$ 
$Z=0.0001$ models. The first line shows our model without
a PMZ, the second line our model with a PMZ of 0.002$\Msun$,
and the third line the yields from the Cristallo et al.
model (C09). We also show the percentage difference between 
line two (PMZ model) and the line three (C09 model).}
\label{m2}
\begin{tabular}{llllllllllll} \hline \hline
Model & \iso{12}C & \iso{14}N & \iso{15}N & \iso{16}O & \iso{19}F & 
\iso{23}Na & \iso{25}Mg & \iso{30}Si & \iso{31}P &  \iso{60}Fe \\  \hline
No PMZ &  3.57($-2$) & 7.08($-5$) & $-3.35(-9)$ & 5.15($-4$) & 1.10($-5$) &  
 1.04($-4$) & 4.87($-5$) & 6.53($-8$) & 1.39($-8$) & 1.18($-8$) \\
PMZ    & 3.27($-2$) & 6.73($-5$) & 1.07($-8$) & 9.52($-4$) & 1.30($-5$) & 
  2.26($-4$) & 7.64($-5$) & 9.14($-8$) & 1.30($-7$) & 7.04($-9$) \\
C09    &  1.71($-2$) & 3.40($-5$) & $-2.07(-8)$ & 4.03($-4$) & 2.44($-6$) & 
  1.38($-5$) & 2.55($-5$) & 9.89($-8$) & 3.53($-7$) & 5.76($-8$) \\ \hline
       & 91.2\% & 98.0\% & 152\% & 136\% & 433\% & 154\% & 200\% & $-$7.60\% &
 $-63.2$\% & $-$87.8\% \\
\hline \hline
\end{tabular}
\medskip\\
\end{center}
\end{table*}

The structural differences between the \citet{cristallo09}
model and ours is reflected in the stellar yields provided 
in Table~\ref{m2}. Cristallo et al. also used different 
initial abundances, scaling their $Z=0.0001$ model to 
the solar abundances of \citet{asplund05} ($Z_{\odot} = 0.0138$, 
smaller than our $Z_{\odot} =0.02$). The differences caused
by this are hard to asses, as they did not provide their 
initial abundances (ours are given in column 7 in the on-line 
yield data tables).  Even so, we find relatively large 
variations for all isotopes, even those not directly produced
by He-burning such as \iso{14}N. 
In these low-metallicity models, most of the \iso{14}N is 
produced during the AGB, as a consequence of the dredge-up of 
material processed by CNO cycling in the H-shell. 
Our yields of He-burning products including \iso{12}C, 
\iso{16}O, \iso{19}F, are larger as expected from our 
model dredging up more He-intershell
material. The increase in the surface abundance of \iso{16}O
results in a final [O/Fe] $\sim 1.0$, and the planetary nebulae
formed from such an object would show the dredge-up of
oxygen. The metallicity dependent enrichment in O predicted
by our models has been reported by \citet{magrini09b} for
planetary nebulae in the galaxy IC~10. 

Back to the comparison to \citet{cristallo09}, while our
yields of He-burning products are larger, the products
of neutron captures (e.g., \iso{31}P) along with \iso{26}Al 
and \iso{27}Al are smaller.  This is a consequence of our
smaller \iso{13}C pocket (at least as measured at 
the 2$^{\rm nd}$ TDU).  The variations range from
only $-14$\% for \iso{26}Al up to $\sim -90$\%
for \iso{60}Fe.  These variations may also reflect 
different choices of reaction rates (e.g., for Al).
One interesting difference is that our
2$\Msun$ model with a PMZ produces a positive yield of 
\iso{15}N, whereas both the model without a PMZ and the
\citet{cristallo09} model have negative yields, indicating
a net destruction. The positive yield indicates that
some \iso{15}N is produced during TPs by the \iso{18}O($p,\alpha$)\iso{15}N
reaction rate, which is part of the fluorine production 
chain \citep[see][]{lugaro04}. The difference in the 
\iso{15}N surface abundance between the models with 
and without a PMZ is $\sim 50$\%, and most of this extra
\iso{15}N is produced during the first few TPs when the
temperature is not yet hot enough for efficient \iso{15}N
destruction (or \iso{19}F production). By the last TPs,
the \iso{15}N produced in the convective TP
is rapidly destroyed in models with and without
a PMZ.

\citet{stancliffe07} made a careful comparison between
the 1.5$\Msun$, $Z =0.008$ yields from \citet{karakas07b}
to their VW93 model, whereas \citet{stancliffe08} 
compared models of $Z=0.0001$. The main difference that
\citet{stancliffe07} found for the 1.5$\Msun$, $Z = 0.008$
model was that their calculation experienced more 
efficient dredge-up, and hence positive
net yields of \iso{12}C, \iso{22}Ne, and \iso{25}Mg.
In comparison, our model had little dredge-up with
only a total of 1.47$\times 10^{-3}\Msun$ of He-shell
material mixed into the envelope. \citet{stancliffe07}
noted that the yields of isotopes dependent on the
efficiency of the first dredge-up (e.g., \iso{14}N)
were reasonably consistent, given that their models
had scaled-solar initial abundances and the
\citet{karakas07b} model did not. From our updated
model with scaled-solar initial abundances, we find 
an agreement of $\sim 5$\% between the \iso{14}N yields.

\citet{stancliffe08} have made
a detailed comparison to the $Z=0.0001$ models available 
in \citet{karakas07b}; here we limit our discussion 
to the new 1.5$\Msun$, $Z=0.0001$ model. Our model
has 18~TPs compared to their 10, and dredged up $\sim 52$\% 
more material from the He-intershell (R. Stancliffe, 
private communication). Hence our yields of
\iso{12}C, \iso{14}N, and \iso{22}Ne are larger by 
9\%, 64\%, and 62\% respectively.  The Stancliffe
\& Glebbeck yields for \iso{23}Na are almost identical,
differing by less than 1\%, whereas their model produces
almost 100\% more \iso{25}Mg and \iso{31}P. 
The last point can be understood by considering that
the \iso{22}Ne($\alpha,n$)\iso{25}Mg reaction is 
marginally activated during TPs at this mass and 
metallicity, and \citet{stancliffe08} used the faster
rate from \citet{kaeppeler94}. 
That the yields for \iso{23}Na are the same is 
a coincidence. Our model produces more \iso{22}Ne 
and would produce more \iso{23}Na, but we use the
new faster \iso{23}Na proton destruction rates.

Lastly, we compare the results from the new 
intermediate-mass $Z=0.0001$ models to the models 
presented by \citet{ventura09}. These authors also 
provide a detailed comparison to \citet{karakas07b} so
here we comment on some of the major changes 
introduced by using the new R75 structure models and
the large reduction in the Na yields. 
The new 4$\Msun$, $Z = 0.0001$ model has a similar
number of TPs (37) to the 4$\Msun$ model computed by
\citet{ventura09} (32), whereas our 6$\Msun$ had
109 TPs compared to their 77 TPs. Furthermore, owing
to the different convective model used in the 
\citet{ventura09} calculations (the Full Spectrum of
Turbulence compared  to our use of the mixing-length
theory with $\alpha = 1.75$), the temperatures at
the base of the envelope in our models were significantly
lower. Our 4$\Msun$ model peaked at 88$\times 10^{6}$K
compared to 92$\times 10^{6}$K; likewise
our 6$\Msun$ peaked at 104$\times 10^{6}$K compared to
127$\times 10^{6}$K in the \citet{ventura09} model.
Finally, one other major difference is evident in
the efficiency of the third dredge-up where our
models show $\lambda \sim 0.9$ for $M \ge 3\Msun$
whereas in the Ventura \& D'Antona models $\lambda$ 
varied from $0.7$ at $4\Msun$ to 0.3 at 6$\Msun$.

These structural differences show up vividly in 
the chemical yields. All our $Z = 0.0001$ models have
a final C/O $> 1$, whereas only the 6$\Msun$, $Z = 0.0001$ 
\citet{ventura09} model has C/O $> 1$ and only as a 
consequence of O destruction. The final surface [O/Fe] 
of our 4$\Msun$ and 6$\Msun$ models are 1.19 and 1.14,
respectively, compared  to 1.08 and $-1.43$ for the
\citet{ventura09} models. The 4$\Msun$ yields for Na
are the same, where the final surface [Na/Fe] abundance 
is 1.89 compared to 1.90 for the \citet{ventura09} 
model. In contrast, the 
[Na/Fe] abundances from the 6$\Msun$ models are
 vastly different: ours is 1.71 compared $-0.20$ 
from \citet{ventura09} model.
This is almost entirely owing to their choice of
convective model that results in much higher HBB 
temperatures and hence higher rates of \iso{23}Na 
destruction.

\section{Conclusions} 

In this paper we present new stellar yields of AGB models
covering a range in initial mass from 1$\Msun$ to 
6$\Msun$, and initial metallicity from $Z=0.02$ to 0.0001.
These yields are an update to the results presented 
in \citet{karakas07b}, using newer reaction rates and 
scaled-solar initial abundances for the 
$Z=0.008$ and 0.004 models. We also present yields covering 
a finer mass grid than used previously, and new 
intermediate-mass $Z=0.0001$ AGB models between 3$\Msun$
and 6$\Msun$. The new $Z=0.0001$ models were evolved
to low ($\sim 0.1\Msun$) envelope masses, with no reduction
in the third dredge-up parameter ($\lambda$) observed.

The main result of this paper is large reductions
in the yields of \iso{23}Na from models with HBB.
This has implications for the chemical evolution of 
galaxies and stellar systems,  and in particular for 
globular clusters. In contrast, the updated reaction 
network only results in small changes to the yields of 
\iso{12}C, \iso{14}N, and \iso{19}F.  The largest 
changes for these isotopes are caused by using new
structure models at $Z= 0.0001$. These new structure
models experience fewer TPs and TDU episodes, and
this results in large reductions in the yields of
\iso{14}N and \iso{17}O.

We examine the effect of a partial mixing 
zone on the stellar yields of low-mass AGB models.
Partial mixing zones are added into models of 2$\Msun$,
$Z = 0.0001$, and 3$\Msun$, $Z = 0.02, 0.008$ and 0.004.
The partial mixing zone results in a \iso{13}C pocket 
in the top $\sim 1/10^{\rm th}$ of the
He-intershell and releases neutrons via the 
\iso{13}C($\alpha,n$)\iso{16}O reaction. The \iso{13}C
pocket not only affects the production of \iso{19}F
but also other isotopes including \iso{23}Na,
the neutron-rich Mg and Si isotopes, \iso{31}P, and \iso{60}Fe.
In particular, the yields of \iso{31}P and \iso{60}Fe
are the most affected by the introduction of the PMZ,
where the yields of \iso{31}P increase by factors of
$\sim 4$ to 20 at 3$\Msun$ depending on metallicity.
Larger variations are  found in the most metal-rich
models.
Because the PMZ is added into the post-processing step
we ignore any feedback onto the stellar structure caused
by the formation of the \iso{13}C pocket. For this
reason the yields should be treated with some
caution. A full detailed study taking into account 
the feedback onto the stellar structure is necessary.

The new results are compared to the yields from 
\citet{karakas07b}, and to other recent AGB yields in 
the literature including \citet{stancliffe07}, 
\citet{cristallo09}, and \citet{ventura09}. The main
result is that the structural differences between the
calculations are still the dominant cause of variations
between the nucleosynthesis predictions. In particular,
the treatment of convection and the algorithm used to
determine convective borders are still important problems
that need to be addressed in the future. 

One important piece of input physics missing from
the models presented here is the inclusion of
carbon and nitrogen-rich low-temperature opacities. While
the approximate treatment of molecular opacities we used
in most models does include some compositional
dependence, the low-temperature tables from \citet{ferguson05}
are for solar or scaled-solar mixtures only. The 
yields presented here and by many other authors show
the complex nucleosynthesis that can occur during the
TP-AGB (e.g., C/O $>1$, C/N $< 1$). The composition
of the envelope needs to be properly taken into account
when determining the opacities of the outer layers. 
Such C and N-rich low-temperature opacity tables are 
now becoming available \citep{lederer09,helling09,marigo09}, 
and used in detailed stellar-structure computations 
\citep{cristallo09}. Stellar models using these 
opacities will need to be computed in the future for 
the size of the mass and metallicity grid 
presented here. 

\section*{Acknowledgments}

The author thanks the referee, Falk Herwig, for critical
comments concerning $^{13}$C pockets in AGB stars.
The author thanks Maria Lugaro for reading the manuscript
and for help with the reaction rate network.
She thanks Simon Campbell for sharing opacity subroutines 
used in the stellar evolution code, and Richard Stancliffe
for providing results ahead of publication.
AIK acknowledges support from the Australian Research 
Council's Discovery Projects funding scheme 
(project number DP0664105).
This study was made possible thanks to the support of the
NCI National Facility at the ANU.

\appendix

\bibliographystyle{mn2e}
\bibliography{mnemonic.bib,/home/akarakas/biblio/library}

\begin{thebibliography}{}

\bibitem[\protect\citeauthoryear{{Anders} \& {Grevesse}}{{Anders} \&
  {Grevesse}}{1989}]{anders89}
{Anders} E.,  {Grevesse} N.,  1989, Geochim. Cosmochim. Acta, 53, 197

\bibitem[\protect\citeauthoryear{{Angulo}, {Arnould}, {Rayet}, {Descouvemont},
  {Baye}, {Leclercq-Willain}, {Coc}, {Barhoumi}, {Aguer}, {Rolfs}, {Kunz},
  {Hammer}, {Mayer}, {Paradellis}, {Kossionides} \& {Chronidou}}{{Angulo}
  et~al.}{1999}]{angulo99}
{Angulo} C., et al.  1999, Nucl. Phys. A, 656, 3

\bibitem[\protect\citeauthoryear{{Asplund}, {Grevesse} \& {Sauval}}{{Asplund}
  et~al.}{2005}]{asplund05}
{Asplund} M.,  {Grevesse} N.,    {Sauval} A.~J.,  2005, in {Barnes} III T.~G.,
  {Bash} F.~N.,  eds, ASP Conf. Ser. 336, ``Cosmic Abundances as Records of
  Stellar Evolution and Nucleosynthesis'', 25

\bibitem[\protect\citeauthoryear{{Bemmerer}, {Confortola}, {Lemut}, {Bonetti},
  {Broggini}, {Corvisiero}, {Costantini}, {Cruz}, {Formicola}, {F{\"u}l{\"o}p},
  {Gervino}, {Guglielmetti} \& {Gustavino}}{{Bemmerer}
  et~al.}{2006}]{bemmerer06}
{Bemmerer} D., et al.  2006, Nucl. Phys. A, 779, 297

\bibitem[\protect\citeauthoryear{{Bessell}, {Brett}, {Wood} \&
  {Scholz}}{{Bessell} et~al.}{1989}]{bessell89}
{Bessell} M.~S.,  {Brett} J.~M.,  {Wood} P.~R.,    {Scholz} M.,  1989, A\&AS,
  77, 1

\bibitem[\protect\citeauthoryear{{Boothroyd} \& {Sackmann}}{{Boothroyd} \&
  {Sackmann}}{1988}]{boothroyd88c}
{Boothroyd} A.~I.,  {Sackmann} I.-J.,  1988, ApJ, 328, 653

\bibitem[\protect\citeauthoryear{{Busso}, {Gallino} \& {Wasserburg}}{{Busso}
  et~al.}{1999}]{busso99}
{Busso} M.,  {Gallino} R.,    {Wasserburg} G.~J.,  1999, ARA\&A, 37, 239

\bibitem[\protect\citeauthoryear{{Campbell} \& {Lattanzio}}{{Campbell} \&
  {Lattanzio}}{2008}]{campbell08}
{Campbell} S.~W.,  {Lattanzio} J.~C.,  2008, A\&A, 490, 769

\bibitem[\protect\citeauthoryear{{Caughlan} \& {Fowler}}{{Caughlan} \&
  {Fowler}}{1988}]{cf88}
{Caughlan} G.~R.,  {Fowler} W.~A.,  1988, Atomic Data and Nuclear Data Tables,
  40, 283

\bibitem[\protect\citeauthoryear{{Chafa}, {Tatischeff}, {Aguer}, {Barhoumi},
  {Coc}, {Garrido}, {Hernanz}, {Jos{\'e}}, {Kiener}, {Lefebvre-Schuhl},
  {Ouichaoui}, {S{\'e}r{\'e}ville} \& {Thibaud}}{{Chafa}
  et~al.}{2007}]{chafa07}
{Chafa} A.,  et al.  2007, Phys. Rev. C, 75, 035810

\bibitem[\protect\citeauthoryear{{Chiosi}, {Wood} \& {Capitanio}}{{Chiosi}
  et~al.}{1993}]{chiosi93}
{Chiosi} C.,  {Wood} P.~R.,    {Capitanio} N.,  1993, ApJS, 86, 541

\bibitem[\protect\citeauthoryear{{Church}, {Cristallo}, {Lattanzio},
  {Stancliffe}, {Straniero} \& {Cannon}}{{Church} et~al.}{2009}]{church09}
{Church} R.~P.,  {Cristallo} S.,  {Lattanzio} J.~C.,  {Stancliffe} R.~J.,
  {Straniero} O.,    {Cannon} R.~C.,  2009, Publications of the Astronomical
  Society of Australia, 26, 217

\bibitem[\protect\citeauthoryear{{Cottrell} \& {Da Costa}}{{Cottrell} \& {Da
  Costa}}{1981}]{cottrell81}
{Cottrell} P.~L.,  {Da Costa} G.~S.,  1981, ApJ, 245, L79

\bibitem[\protect\citeauthoryear{{Cristallo}, {Straniero}, {Gallino},
  {Piersanti}, {Dom{\'{\i}}nguez} \& {Lederer}}{{Cristallo}
  et~al.}{2009}]{cristallo09}
{Cristallo} S.,  {Straniero} O.,  {Gallino} R.,  {Piersanti} L.,
  {Dom{\'{\i}}nguez} I.,    {Lederer} M.~T.,  2009, ApJ, 696, 797

\bibitem[\protect\citeauthoryear{{Decressin}, {Charbonnel}, {Siess} \&
  {Palacios}}{{Decressin} et~al.}{2004}]{decressin04}
{Decressin} T.,  {Charbonnel} C.,  {Siess} L.,    {Palacios} A.,  2004, Memorie
  della Societa Astronomica Italiana, 75, 682

\bibitem[\protect\citeauthoryear{{Denker}, {Drotleff}, {Grosse}, {Knee},
  {Kunz}, {Mayer}, {Seidel}, {Soin{\'e}}, {W{\"o}ohr}, {Wolf} \&
  {Hammer}}{{Denker} et~al.}{1995}]{denker95}
{Denker} A., et al.  eds, Nuclei in the Cosmos III Vol.~327, 255

\bibitem[\protect\citeauthoryear{{El Eid} \& {Champagne}}{{El Eid} \&
  {Champagne}}{1995}]{eleid95}
{El Eid} M.~F.,  {Champagne} A.~E.,  1995, ApJ, 451, 298

\bibitem[\protect\citeauthoryear{{Fenner}, {Campbell}, {Karakas}, {Lattanzio}
  \& {Gibson}}{{Fenner} et~al.}{2004}]{fenner04}
{Fenner} Y.,  {Campbell} S.,  {Karakas} A.~I.,  {Lattanzio} J.~C.,    {Gibson}
  B.~K.,  2004, MNRAS, 353, 789

\bibitem[\protect\citeauthoryear{{Ferguson}, {Alexander}, {Allard}, {Barman},
  {Bodnarik}, {Hauschildt}, {Heffner-Wong} \& {Tamanai}}{{Ferguson}
  et~al.}{2005}]{ferguson05}
{Ferguson} J.~W.,  et al.  2005, ApJ, 623, 585

\bibitem[\protect\citeauthoryear{{Forestini}, {Goriely}, {Jorissen} \&
  {Arnould}}{{Forestini} et~al.}{1992}]{forestini92}
{Forestini} M.,  {Goriely} S.,  {Jorissen} A.,    {Arnould} M.,  1992, A\&A,
  261, 157

\bibitem[\protect\citeauthoryear{{Frost} \& {Lattanzio}}{{Frost} \&
  {Lattanzio}}{1996}]{frost96}
{Frost} C.~A.,  {Lattanzio} J.~C.,  1996, ApJ, 473, 383

\bibitem[\protect\citeauthoryear{{Fynbo}, {Diget}, {Bergmann}, {Borge},
  {Cederk{\"a}ll}, {Dendooven}, {Fraile}, {Franchoo} \& {Fedosseev}}{{Fynbo}
  et~al.}{2005}]{fynbo05}
{Fynbo} H.~O.~U.,  et al. 2005, Nature, 433, 136

\bibitem[\protect\citeauthoryear{{Gallino}, {Arlandini}, {Busso}, {Lugaro},
  {Travaglio}, {Straniero}, {Chieffi} \& {Limongi}}{{Gallino}
  et~al.}{1998}]{gallino98}
{Gallino} R., et al. 1998, ApJ, 497, 388

\bibitem[\protect\citeauthoryear{{Goriely} \& {Mowlavi}}{{Goriely} \&
  {Mowlavi}}{2000}]{goriely00}
{Goriely} S.,  {Mowlavi} N.,  2000, A\&A, 362, 599

\bibitem[\protect\citeauthoryear{{Goriely} \& {Siess}}{{Goriely} \&
  {Siess}}{2004}]{goriely04}
{Goriely} S.,  {Siess} L.,  2004, A\&A, 421, L25

\bibitem[\protect\citeauthoryear{{Gratton}, {Sneden} \& {Carretta}}{{Gratton}
  et~al.}{2004}]{gratton04}
{Gratton} R.,  {Sneden} C.,    {Carretta} E.,  2004, ARA\&A, 42, 385

\bibitem[\protect\citeauthoryear{{Hale}, {Champagne}, {Iliadis}, {Hansper},
  {Powell} \& {Blackmon}}{{Hale} et~al.}{2002}]{hale02}
{Hale} S.~E.,  {Champagne} A.~E.,  {Iliadis} C.,  {Hansper} V.~Y.,  {Powell}
  D.~C.,    {Blackmon} J.~C.,  2002, Phys. Rev. C, 65, 015801

\bibitem[\protect\citeauthoryear{{Hale}, {Champagne}, {Iliadis}, {Hansper},
  {Powell} \& {Blackmon}}{{Hale} et~al.}{2004}]{hale04}
{Hale} S.~E.,  {Champagne} A.~E.,  {Iliadis} C.,  {Hansper} V.~Y.,  {Powell}
  D.~C.,    {Blackmon} J.~C.,  2004, Phys. Rev. C, 70, 045802

\bibitem[\protect\citeauthoryear{{Helling} \& {Lucas}}{{Helling} \&
  {Lucas}}{2009}]{helling09}
{Helling} C.,  {Lucas} W.,  2009, MNRAS, 398, 985

\bibitem[\protect\citeauthoryear{{Herwig}}{{Herwig}}{2004a}]{herwig04a}
{Herwig} F.,  2004a, ApJ, 605, 425

\bibitem[\protect\citeauthoryear{{Herwig}}{{Herwig}}{2004b}]{herwig04b}
{Herwig} F.,  2004b, ApJS, 155, 651

\bibitem[\protect\citeauthoryear{{Herwig}}{{Herwig}}{2005}]{herwig05}
{Herwig} F.,  2005, ARA\&A, 43, 435

\bibitem[\protect\citeauthoryear{{Iglesias} \& {Rogers}}{{Iglesias} \&
  {Rogers}}{1996}]{iglesias96}
{Iglesias} C.~A.,  {Rogers} F.~J.,  1996, ApJ, 464, 943

\bibitem[\protect\citeauthoryear{{Iliadis}, {D'Auria}, {Starrfield}, {Thompson}
  \& {Wiescher}}{{Iliadis} et~al.}{2001}]{iliadis01}
{Iliadis} C.,  {D'Auria} J.~M.,  {Starrfield} S.,  {Thompson} W.~J.,
  {Wiescher} M.,  2001, ApJS, 134, 151

\bibitem[\protect\citeauthoryear{{Izzard}, {Lugaro}, {Karakas}, {Iliadis} \&
  {van Raai}}{{Izzard} et~al.}{2007}]{izzard07}
{Izzard} R.~G.,  {Lugaro} M.,  {Karakas} A.~I.,  {Iliadis} C.,    {van Raai}
  M.,  2007, A\&A, 466, 641

\bibitem[\protect\citeauthoryear{{Izzard}, {Tout}, {Karakas} \&
  {Pols}}{{Izzard} et~al.}{2004}]{izzard04b}
{Izzard} R.~G.,  {Tout} C.~A.,  {Karakas} A.~I.,    {Pols} O.~R.,  2004, MNRAS,
  350, 407

\bibitem[\protect\citeauthoryear{{Kaeppeler}, {Wiescher}, {Giesen}, {Goerres},
  {Baraffe}, {El Eid}, {Raiteri}, {Busso}, {Gallino}, {Limongi} \&
  {Chieffi}}{{Kaeppeler} et~al.}{1994}]{kaeppeler94}
{Kaeppeler} F., et al. 1994, ApJ, 437, 396

\bibitem[\protect\citeauthoryear{{Karakas} \& {Lattanzio}}{{Karakas} \&
  {Lattanzio}}{2003}]{karakas03a}
{Karakas} A.~I.,  {Lattanzio} J.~C.,  2003, PASA, 20, 393

\bibitem[\protect\citeauthoryear{{Karakas} \& {Lattanzio}}{{Karakas} \&
  {Lattanzio}}{2007}]{karakas07b}
{Karakas} A.~I.,  {Lattanzio} J.~C.,  2007, PASA, 24, 103

\bibitem[\protect\citeauthoryear{{Karakas}, {Lattanzio} \& {Pols}}{{Karakas}
  et~al.}{2002}]{karakas02}
{Karakas} A.~I.,  {Lattanzio} J.~C.,    {Pols} O.~R.,  2002, PASA, 19, 515

\bibitem[\protect\citeauthoryear{{Karakas}, {Lee}, {Lugaro}, {G{\"o}rres} \&
  {Wiescher}}{{Karakas} et~al.}{2008}]{karakas08}
{Karakas} A.~I.,  {Lee} H.~Y.,  {Lugaro} M.,  {G{\"o}rres} J.,    {Wiescher}
  M.,  2008, ApJ, 676, 1254

\bibitem[\protect\citeauthoryear{{Karakas}, {Lugaro}, {Wiescher}, {Goerres} \&
  {Ugalde}}{{Karakas} et~al.}{2006}]{karakas06a}
{Karakas} A.~I.,  {Lugaro} M.,  {Wiescher} M.,  {Goerres} J.,    {Ugalde} C.,
  2006, ApJ, 643, 471

\bibitem[\protect\citeauthoryear{{Karakas}, {van Raai}, {Lugaro}, {Sterling} \&
  {Dinerstein}}{{Karakas} et~al.}{2009}]{karakas09}
{Karakas} A.~I.,  {van Raai} M.~A.,  {Lugaro} M.,  {Sterling} N.~C.,
  {Dinerstein} H.~L.,  2009, ApJ, 690, 1130

\bibitem[\protect\citeauthoryear{{Kroupa}, {Tout} \& {Gilmore}}{{Kroupa}
  et~al.}{1993}]{kroupa93}
{Kroupa} P.,  {Tout} C.~A.,    {Gilmore} G.,  1993, MNRAS, 262, 545

\bibitem[\protect\citeauthoryear{{Lederer} \& {Aringer}}{{Lederer} \&
  {Aringer}}{2009}]{lederer09}
{Lederer} M.~T.,  {Aringer} B.,  2009, A\&A, 494, 403

\bibitem[\protect\citeauthoryear{{Lee}}{{Lee}}{2006}]{LeeThesis}
{Lee} H.~Y.,  2006, PhD thesis, University of Notre Dame

\bibitem[\protect\citeauthoryear{{Lugaro}, {Ugalde}, {Karakas}, {G{\"o}rres},
  {Wiescher}, {Lattanzio} \& {Cannon}}{{Lugaro} et~al.}{2004}]{lugaro04}
{Lugaro} M.,  {Ugalde} C.,  {Karakas} A.~I.,  {G{\"o}rres} J.,  {Wiescher} M.,
  {Lattanzio} J.~C.,    {Cannon} R.~C.,  2004, ApJ, 615, 934

\bibitem[\protect\citeauthoryear{{Magrini} \& {Gon{\c c}alves}}{{Magrini} \&
  {Gon{\c c}alves}}{2009}]{magrini09b}
{Magrini} L.,  {Gon{\c c}alves} D.~R.,  2009, MNRAS, 398, 280

\bibitem[\protect\citeauthoryear{{Marigo}}{{Marigo}}{2001}]{marigo01}
{Marigo} P.,  2001, A\&A, 370, 194

\bibitem[\protect\citeauthoryear{{Marigo}}{{Marigo}}{2002}]{marigo02}
{Marigo} P.,  2002, A\&A, 387, 507

\bibitem[\protect\citeauthoryear{{Marigo} \& {Aringer}}{{Marigo} \&
  {Aringer}}{2009}]{marigo09}
{Marigo} P.,  {Aringer} B.,  2009, A\&A, submitted

\bibitem[\protect\citeauthoryear{{Reimers}}{{Reimers}}{1975}]{reimers75}
{Reimers} D.,  1975, {Circumstellar envelopes and mass loss of red giant
  stars}.
Problems in stellar atmospheres and envelopes., pp 229--256

\bibitem[\protect\citeauthoryear{{Renzini}}{{Renzini}}{2008}]{renzini08}
{Renzini} A.,  2008, MNRAS, 391, 354

\bibitem[\protect\citeauthoryear{{Sakharuk}, {Elliot}, {Fisker}, {Hemingray},
  {Kruizenga}, {Rauscher}, {Schatz}, {Smith}, {Thielemann} \&
  {Wiescher}}{{Sakharuk} et~al.}{2006}]{sakharuk06}
{Sakharuk} A., et al., ``Capture Gamma-Ray Spectroscopy and Related
Topics Vol.~819 of American Institute of Physics Conf. Series'', 118

\bibitem[\protect\citeauthoryear{{Smith}, {Lambert} \& {McWilliam}}{{Smith}
  et~al.}{1987}]{smith87}
{Smith} V.~V.,  {Lambert} D.~L.,    {McWilliam} A.,  1987, ApJ, 320, 862

\bibitem[\protect\citeauthoryear{{Stancliffe} \& {Glebbeek}}{{Stancliffe} \&
  {Glebbeek}}{2008}]{stancliffe08}
{Stancliffe} R.~J.,  {Glebbeek} E.,  2008, MNRAS, 389, 1828

\bibitem[\protect\citeauthoryear{{Stancliffe} \& {Jeffery}}{{Stancliffe} \&
  {Jeffery}}{2007}]{stancliffe07}
{Stancliffe} R.~J.,  {Jeffery} C.~S.,  2007, MNRAS, 375, 1280

\bibitem[\protect\citeauthoryear{{Straniero}, {Chieffi}, {Limongi}, {Busso},
  {Gallino} \& {Arlandini}}{{Straniero} et~al.}{1997}]{straniero97}
{Straniero} O.,  {Chieffi} A.,  {Limongi} M.,  {Busso} M.,  {Gallino} R.,
  {Arlandini} C.,  1997, ApJ, 478, 332

\bibitem[\protect\citeauthoryear{{Straniero}, {Gallino}, {Busso}, {Chiefei},
  {Raiteri}, {Limongi} \& {Salaris}}{{Straniero} et~al.}{1995}]{straniero95}
{Straniero} O.,  {Gallino} R.,  {Busso} M.,  {Chiefei} A.,  {Raiteri} C.~M.,
  {Limongi} M.,    {Salaris} M.,  1995, ApJ, 440, L85

\bibitem[\protect\citeauthoryear{{Ugalde}, {Azuma}, {Couture}, {G{\"o}rres},
  {Lee}, {Stech}, {Strandberg}, {Tan} \& {Wiescher}}{{Ugalde}
  et~al.}{2008}]{ugalde08}
{Ugalde} C.,  et al.  2008, Phys. Rev. C, 77, 035801

\bibitem[\protect\citeauthoryear{{Vassiliadis} \& {Wood}}{{Vassiliadis} \&
  {Wood}}{1993}]{vw93}
{Vassiliadis} E.,  {Wood} P.~R.,  1993, ApJ, 413, 641

\bibitem[\protect\citeauthoryear{{Ventura} \& {D'Antona}}{{Ventura} \&
  {D'Antona}}{2005}]{ventura05b}
{Ventura} P.,  {D'Antona} F.,  2005, A\&A, 439, 1075

\bibitem[\protect\citeauthoryear{{Ventura} \& {D'Antona}}{{Ventura} \&
  {D'Antona}}{2009}]{ventura09}
{Ventura} P.,  {D'Antona} F.,  2009, A\&A, 499, 835

\bibitem[\protect\citeauthoryear{{Woodward}, {Herwig}, {Porter}, {Fuchs},
  {Nowatzki} \& {Pignatari}}{{Woodward} et~al.}{2008}]{woodward08}
{Woodward} P.,  {Herwig} F.,  {Porter} D.,  {Fuchs} T.,  {Nowatzki} A.,
  {Pignatari} M.,  2008, in {O'Shea} B.~W.,  {Heger} A.,  eds, ``First Stars III'',
  Vol.~990 of American Institute of Physics Conference Series, 300


\end{thebibliography}

\appendix

\section[]{Examples of the on-line tables}

In Table~\ref{example1} we show the first few lines of 
the on-line table containing structural information
from the new AGB models.
In Tables~\ref{table-yieldz02},~\ref{table-yieldz008},
\ref{table-yieldz004}, \ref{table-yieldz0001} we show 
the first few lines of the on-line stellar yield tables for the 
$Z = 0.02$, $Z= 0.008$, $Z=0.004$, and $Z=0.0001$ models,
respectively. An example of the on-line stellar yield 
for the models with partial mixing zones is shown 
in  Table~\ref{table-yieldpmz}.

\clearpage

\begin{table}
\begin{minipage}[]{120mm}
\begin{center}
\caption{The first few lines from Table~A1. Each model entry starts
with a header providing the initial mass and metallicity of the
model. To fit onto the page, we only show the first 8 columns.}
\label{example1}
\begin{tabular}{cccccccccccccc}
\multicolumn{8}{l}{\#Minitial =  4.50 msun, Z = 0.0200}\\
\hline \#pulse & $M_{\rm core}$ & $M_{\rm csh}$ & $t_{\rm csh}$ &
 $\Delta M_{\rm dredge}$ & $\lambda$ & $\lambda_{\rm dup}$ & $T_{\rm
Heshell}$ \\
\hline 
\hline
1 & 8.395468E-01 & 2.890170E-03 & 4.988529E+01 & 0.000000E+00 & 
0.000000E+00 & 0.000000E+00 & 2.125615E+08 \\
2 & 8.407463E-01 & 4.578114E-03 & 3.937420E+01 & 0.000000E+00 & 
0.000000E+00 & 0.000000E+00 & 2.376553E+08 \\
3 & 8.421246E-01 & 4.609287E-03 & 3.606038E+01 & 1.928210E-04 & 
1.426807E-01 & 4.183316E-02 & 2.482608E+08 \\
\hline
\hline
\end{tabular}
\medskip\\
\end{center}
\end{minipage}
\end{table}

\begin{table}
\begin{minipage}[]{120mm}
\begin{center}
\caption{The first few lines from the Table~A2. Each model entry
begins with a header providing the initial mass and final mass
(in solar units) along with the metallicity.
}\label{table-yieldz02}
\begin{tabular}{ccrrrrrr}
\multicolumn{8}{l}{\# Minitial =  1.00 msun, Z = 0.0200, Mfinal =
0.564 msun} \\
\hline Isotope $i$ &  $A$ & yield & mass$(i)_{\rm lost}$ &
mass$(i)_{0}$ & $\langle X(i) \rangle$ & $X0(i)$ & $f$ \\
\hline
\hline
$g^{a}$   & 1 & 8.1815390E-09 & 3.2919932E-08 & 2.4738393E-08 & 
7.5504431E-08 & 5.6745975E-08 & 1.2403737E-01 \\
n         & 1 & -3.9796876E-43 & -3.9796876E-43 & 0.0000000E+00 &
-9.1224530E-43 & 0.0000000E+00 & 0.0000000E+00  \\
p 	  & 1 & -8.1048310E-03 & 2.9179156E-01 & 2.9989639E-01 & 
6.6924673E-01 & 6.8791503E-01 & -1.1948545E-02 \\
\hline
\hline
\end{tabular}
\medskip\\
$^a$ $g$ represents the sum of abundances from \iso{64}Ni to Bi; an increase in $g$ 
indicates that neutron-captures have occurred beyond the end of the
network.\\
\end{center}
\end{minipage}
\end{table}

\begin{table}
\begin{minipage}[]{120mm}
\begin{center}
\caption{The first few lines from the Table~A3. 
}\label{table-yieldz008}
\begin{tabular}{ccrrrrrr}
\multicolumn{8}{l}{\# Minitial =  1.00 msun, Z = 0.0080, Mfinal =
0.577 msun} \\
\hline Isotope $i$ &  $A$ & yield & mass$(i)_{\rm lost}$ &
mass$(i)_{0}$ & $\langle X(i) \rangle$ & $X0(i)$ & $f$ \\
\hline
\hline
g         & 1 & 3.3505394E-09 & 1.2944828E-08 & 9.5942889E-09 & 
3.0602433E-08 & 2.2684230E-08 & 1.3003190E-01 \\
n         & 1 & 0.0000000E+00 & 0.0000000E+00 & 0.0000000E+00 & 
0.0000000E+00 & 0.0000000E+00 & 0.0000000E+00 \\
p 	  & 1 & -8.7296963E-03 & 3.0307159E-01 & 3.1180128E-01 & 
7.1648133E-01 & 7.3720652E-01 & -1.2384283E-02 \\
\hline
\hline
\end{tabular}
\medskip\\
\end{center}
\end{minipage}
\end{table}

\begin{table}
\begin{minipage}[]{120mm}
\begin{center}
\caption{The first few lines from the Table~A4.
}\label{table-yieldz004}
\begin{tabular}{ccrrrrrr}
\multicolumn{8}{l}{\# Minitial =  1.00 msun, Z = 0.0040, Mfinal =
0.610 msun} \\
\hline Isotope $i$ &  $A$ & yield & mass$(i)_{\rm lost}$ &
mass$(i)_{0}$ & $\langle X(i) \rangle$ & $X0(i)$ & $f$ \\
\hline
\hline
g         & 1 & 1.6834560E-09 & 6.1053904E-09 & 4.4219344E-09 & 
1.5654848E-08 & 1.1339757E-08 & 1.4004514E-01 \\
n         & 1 & 0.0000000E+00 & 0.0000000E+00 & 0.0000000E+00 & 
0.0000000E+00 & 0.0000000E+00 & 0.0000000E+00 \\
p 	  & 1 & -8.1304908E-03 & 2.8376853E-01 & 2.9189903E-01 & 
7.2761166E-01 & 7.4855560E-01 & -1.2324413E-02 \\
\hline
\hline
\end{tabular}
\medskip\\
\end{center}
\end{minipage}
\end{table}

\begin{table}
\begin{minipage}[]{120mm}
\begin{center}
\caption{The first few lines from the Table~A5.
}\label{table-yieldz0001}
\begin{tabular}{ccrrrrrr}
\multicolumn{8}{l}{\# Minitial =  1.00 msun, Z = 0.0001, Mfinal =
0.720 msun} \\
\hline Isotope $i$ &  $A$ & yield & mass$(i)_{\rm lost}$ &
mass$(i)_{0}$ & $\langle X(i) \rangle$ & $X0(i)$ & $f$ \\
\hline
\hline
g         & 1 & 5.8190168E-09 & 5.8983649E-09 & 7.9347945E-11 & 
2.1065592E-08 & 2.8343647E-10 & 1.8711179E+00 \\
n         & 1 & -3.2538150E-42 & -3.2538150E-42 & 0.0000000E+00 & 
-1.1620968E-41 & 0.0000000E+00 & 0.0000000E+00 \\
p 	  & 1 & -5.6426227E-03 & 2.0989175E-01 & 2.1553437E-01 & 
7.4961346E-01 & 7.6990402E-01 & -1.1599216E-02 \\
\hline
\hline
\end{tabular}
\medskip\\
\end{center}
\end{minipage}
\end{table}

\clearpage

\begin{table}
\begin{minipage}[]{120mm}
\begin{center}
\caption{The first few lines from the Table~A6.
}\label{table-yieldpmz}
\begin{tabular}{ccrrrrrr}
\multicolumn{8}{l}{\# Minitial =  3.00 msun, Z = 0.0200, Mfinal =
0.682 msun, partial mixing zone = 2e-3 msun} \\
\hline Isotope $i$ &  $A$ & yield & mass$(i)_{\rm lost}$ &
mass$(i)_{0}$ & $\langle X(i) \rangle$ & $X0(i)$ & $f$ \\
\hline
\hline
g         & 1 & 3.4544257E-06 & 3.5859543E-06 & 1.3152861E-07 & 
1.5470035E-06 & 5.6745975E-08 & 1.4355562E+00 \\
n         & 1 & 0.0000000E+00 & 0.0000000E+00 & 0.0000000E+00 & 
0.0000000E+00 & 0.0000000E+00 & 0.0000000E+00 \\
p 	  & 1 & -6.9546103E-02 & 1.5249372E+00 & 1.5944833E+00 & 
6.5786761E-01 & 6.8791503E-01 & -1.9396281E-02 \\
\hline
\hline
\end{tabular}
\medskip\\
\end{center}
\end{minipage}
\end{table}

\bsp

\label{lastpage}

\end{document}